\documentclass[envcountsame,a4paper]{llncs}
\usepackage{ifthen}
\PassOptionsToPackage{dvipsnames}{xcolor}
\usepackage{xcolor}
\usepackage[noend]{algpseudocode}
\usepackage{algorithm}
\usepackage{amsfonts}
\usepackage{amsmath}

\usepackage{amssymb}
\usepackage{array}
\usepackage[english]{babel}
\usepackage{booktabs}
\usepackage{braket}
\usepackage{caption}
\usepackage[noadjust]{cite}
\usepackage{csquotes}
\usepackage{eso-pic}
\usepackage{graphicx}
\PassOptionsToPackage{hidelinks,colorlinks=true,linkcolor=Blue,urlcolor=Blue,citecolor=OliveGreen,bookmarksdepth=3,bookmarksopen=true}{hyperref}
\usepackage{hyperref}
\usepackage[misc]{ifsym}
\usepackage[utf8]{inputenc}
\usepackage{lipsum}
\usepackage{marginnote}
\usepackage{mathtools}
\usepackage{mwe}
\usepackage{pgfplots}
\usepackage{subfig}
\usepackage{tabularx}
\usepackage{tikz}
\usepackage{xparse}
\usepackage{xspace}
\usepackage{xurl}

\newif\iffullversion 
\fullversiontrue

\newif\ifdraft 
\drafttrue

\newtheorem{experiment}[theorem]{Experiment}

\newtheorem{dfn}[theorem]{Definition}

\ifdraft

\newcommand{\old}[1]{{\color{RawSienna}\sout{#1}}} 
\newcommand{\tg}[1]{{\color{red}{\textsf{\textbf{TG:} #1}}}} 
\else

\newcommand{\old}[1]{}
\newcommand{\tg}[1]{}
\fi

\newcommand{\thisdate}{2024-08-23\xspace}
\newcommand{\sflclastversion}{v0.4.5\xspace}
\newcommand{\sflcbenchmarkversion}{v0.4.1\xspace}
\newcommand{\osbenchmark}{Ubuntu 23.04\xspace}
\newcommand{\kernelbenchmark}{v6.2.0\xspace}
\newcommand{\luksbenchmarkversion}{v6.2.0-26\xspace}
\newcommand{\veracryptbenchmarkversion}{v1.25.9\xspace}
\newcommand{\dmsflc}{\ensuremath{\mathtt{dm\mbox{-}sflc}}\xspace}
\newcommand{\shufflecake}{\ensuremath{\mathtt{shufflecake}}\xspace}

\newcommand{\expref}[2]{\texorpdfstring{\hyperref[#2]{#1~\ref{#2}}}{#1~\ref{#2}}}
\renewcommand{\epsilon}{\ensuremath{\varepsilon}\xspace}
\renewcommand{\phi}{\ensuremath{\varphi}\xspace}
\renewcommand{\Psi}{\ensuremath{\varPsi}\xspace}
\newcommand{\from}{\ensuremath{\leftarrow}\xspace}

\renewcommand{\Pr}[2][]{\ensuremath{ \underset{#1}{\mathsf{Pr}} \left[ {#2} \right] \xspace}} 
\newcommand{\fromunif}{\ensuremath{\raisebox{-1pt}{\ensuremath{\,\xleftarrow{\raisebox{-1pt}{$\scriptscriptstyle\$$}}\,}}}} 
\newcommand{\negl}{\ensuremath{\mathsf{negl}}\xspace} 

\newcommand{\secpar}{\ensuremath{\lambda}\xspace}

\newcommand{\free}{\ensuremath{\mathsf{free}}\xspace}
\newcommand{\occupied}{\ensuremath{\mathsf{occupied}}\xspace}

\newcommand{\key}{\ensuremath{{k}}\xspace}

\newcommand{\ptxt}{\ensuremath{{p}}\xspace}
\newcommand{\ctxt}{\ensuremath{{c}}\xspace}
\newcommand{\rnd}{\ensuremath{{r}}\xspace}

\newcommand{\salt}{\ensuremath{\mathsf{salt}}\xspace}

\newcommand{\IV}{\ensuremath{\mathsf{IV}}\xspace}

\newcommand{\adver}{\ensuremath{\mathcal{A}}\xspace}

\newcommand{\escheme}{\ensuremath{\mathcal{E}}\xspace}

\newcommand{\Decrypt}{\ensuremath{\mathtt{Decrypt}}\xspace}

\newcommand{\Encrypt}{\ensuremath{\mathtt{Encrypt}}\xspace}

\newcommand{\KDF}{\ensuremath{\mathtt{KDF}}\xspace}

\newcommand{\win}{\ensuremath{\mathsf{win}}\xspace}

\newcommand{\advantage}[3]{\ensuremath{\mathbf{Adv}_{#1}^{#2}\left(#3\right)}\xspace} 

\newcommand{\INDCPA}{\ensuremath{\mathsf{IND{\mbox{-}}CPA}}\xspace}

\newcommand{\oracle}{\ensuremath{\mathcal{O}}\xspace}

\newcommand{\nread}{\ensuremath{\mathtt{read}}\xspace}
\newcommand{\nwrite}{\ensuremath{\mathtt{write}}\xspace}
\newcommand{\bRead}{\ensuremath{\mathtt{bRead}}\xspace}
\newcommand{\bWrite}{\ensuremath{\mathtt{bWrite}}\xspace}
\newcommand{\SflcRead}{\ensuremath{\mathtt{SflcRead}}\xspace}
\newcommand{\SflcWrite}{\ensuremath{\mathtt{SflcWrite}}\xspace}
\newcommand{\NewSlice}{\ensuremath{\mathtt{NewSlice}}\xspace}
\newcommand{\Instantiate}{\ensuremath{\mathtt{Instantiate}}\xspace}
\newcommand{\sflcinit}{\ensuremath{\mathtt{init}}\xspace}
\newcommand{\sflcopen}{\ensuremath{\mathtt{open}}\xspace}
\newcommand{\sflcclose}{\ensuremath{\mathtt{close}}\xspace}
\newcommand{\HandleCorruption}{\ensuremath{\mathtt{HandleCorruption}}\xspace}
\newcommand{\ReclaimSlice}{\ensuremath{\mathtt{ReclaimSlice}}\xspace}
\newcommand{\LoadIV}{\ensuremath{\mathtt{LoadIV}}\xspace}
\newcommand{\SampleAndStoreIV}{\ensuremath{\mathtt{SampleAndStoreIV}}\xspace}
\newcommand{\ReverseShuffle}{\ensuremath{\mathtt{ReverseShuffle}}\xspace}

\newcommand{\block}{\ensuremath{B}\xspace}
\newcommand{\blocksize}{\ensuremath{{S_\mathsf{B}}}\xspace}
\newcommand{\keysize}{\ensuremath{{S_\mathsf{K}}}\xspace}
\newcommand{\devbsize}{\ensuremath{{S_\mathsf{D}}}\xspace}
\newcommand{\Nvol}{\ensuremath{N}\xspace}
\newcommand{\maxvols}{\ensuremath{N_\mathsf{MAX}}\xspace}
\newcommand{\Sl}{\ensuremath{S_\mathsf{L}}\xspace}
\newcommand{\Sp}{\ensuremath{S_\mathsf{P}}\xspace}
\newcommand{\Shdr}{\ensuremath{S_\mathsf{H}}\xspace}
\newcommand{\deltas}{\ensuremath{\Delta_\mathsf{S}}\xspace}
\newcommand{\PSI}{\ensuremath{\psi}\xspace}
\newcommand{\LSI}{\ensuremath{\sigma}\xspace}
\newcommand{\SliceMap}{\ensuremath{\mathsf{SliceMap}}\xspace}
\newcommand{\numslices}{\ensuremath{\mathsf{numslices}}\xspace}
\newcommand{\maxslices}{\ensuremath{\mathsf{mxslc}}\xspace}
\newcommand{\V}{\ensuremath{V}\xspace}
\newcommand{\Blog}{\ensuremath{\beta_\mathsf{L}}\xspace}
\newcommand{\Bphy}{\ensuremath{\beta_\mathsf{P}}\xspace}
\newcommand{\prmslices}{\ensuremath{\mathsf{prmslices}}\xspace}
\newcommand{\ofld}{\ensuremath{\mathsf{ofld}}\xspace}
\newcommand{\octr}{\ensuremath{\mathsf{octr}}\xspace}
\newcommand{\KEK}{\ensuremath{\mathsf{KEK}}\xspace}
\newcommand{\VMK}{\ensuremath{\mathsf{VMK}}\xspace}
\newcommand{\VEK}{\ensuremath{\mathsf{VEK}}\xspace}
\newcommand{\metadata}{\ensuremath{\mathsf{metadata}}\xspace}
\newcommand{\dvc}{\ensuremath{\mathsf{device}}}
\newcommand{\pwd}{\ensuremath{\mathsf{password}}}
\newcommand{\DMB}{\ensuremath{\mathsf{DMB}}}
\newcommand{\cell}{\ensuremath{\mathsf{cell}}}
\newcommand{\VMB}{\ensuremath{\mathsf{VMB}}}

\begin{document}


\title{Shufflecake: Plausible Deniability for Multiple Hidden Filesystems on Linux}
\author{Elia Anzuoni\inst{1, 2} \and
Tommaso Gagliardoni\inst{2} \Letter}
\institute{EPFL, Switzerland \\ \email{elianzuoni@gmaildotcom} 
\and
Kudelski Security, Switzerland \\ \email{myfirstname@gagliardoni.net}
}

\let\oldaddcontentsline\addcontentsline
\def\addcontentsline#1#2#3{}
\maketitle
\def\addcontentsline#1#2#3{\oldaddcontentsline{#1}{#2}{#3}}
\noindent

\iffullversion
	\AddToShipoutPicture*
	{
	\centering\small
	\raisebox{5.0cm}
	{
	\hspace{4.5cm}\parbox{\textwidth}
	{
		\begin{small}
		\begin{center}
		A 15-page abstract of this work appears (with the same title) in the proceedings of the \emph{ACM Conference on Computer and Communications Security (CCS) 2023}. This is the authors' full version. This document supersedes any previous versions.\\ \ \\Original Version: 2023-10-06. \ \ \ \ \ This Revision: \thisdate. \ \ \ \ \ \\ Reference Software Implementation Version: \sflclastversion (`Legacy' scheme only)
		\end{center}
		\end{small}
	}
	}
	}
\fi

\begin{abstract}
We present Shufflecake, a new plausible deniability design to hide the existence of encrypted data on a storage medium making it very difficult for an adversary to prove the existence of such data. Shufflecake can be considered a ``spiritual successor'' of tools such as TrueCrypt and VeraCrypt, but vastly improved: it works natively on Linux, it supports any filesystem of choice, and can manage multiple volumes per device, so to make deniability of the existence of hidden partitions really plausible. 

Compared to ORAM-based solutions, Shufflecake is extremely fast and simpler but does not offer native protection against multi-snapshot adversaries. However, we discuss security extensions that are made possible by its architecture, and we show evidence why these extensions might be enough to thwart more powerful adversaries. 

We implemented Shufflecake as an in-kernel tool for Linux, adding useful features, and we benchmarked its performance showing only a minor slowdown compared to a base encrypted system. We believe Shufflecake represents a useful tool for people whose freedom of expression is threatened by repressive authorities or dangerous criminal organizations, in particular: whistleblowers, investigative journalists, and activists for human rights in oppressive regimes.


\vspace{0.8cm}
\begin{figure*}[ht]
\centering
\includegraphics[width=0.40\textwidth]{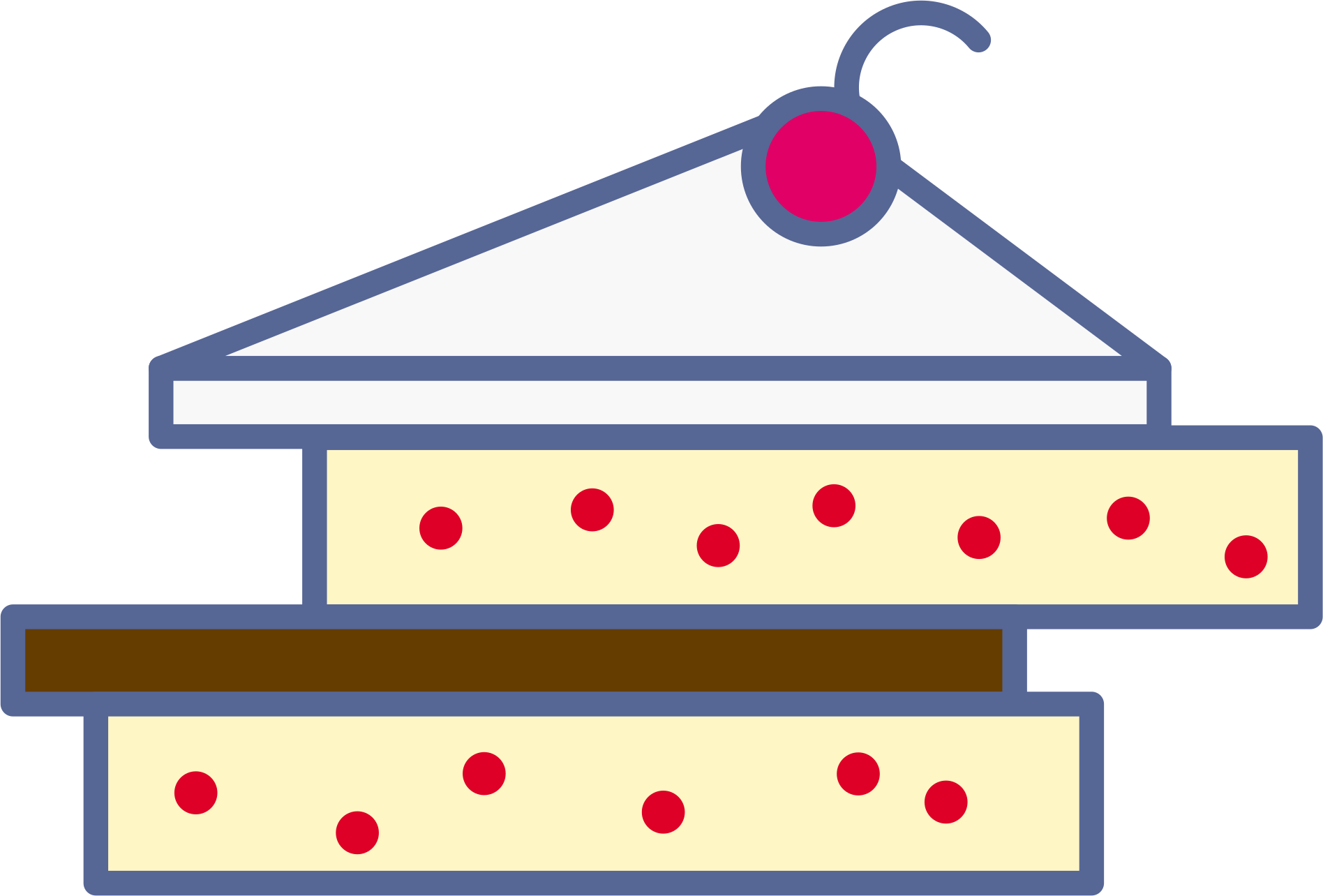}
\end{figure*}

\end{abstract}

\setcounter{tocdepth}{2}
\tableofcontents
\newpage


\section{Introduction}\label{sec:introduction}

Privacy of personal and sensitive data is, now more than ever, a topic of major public interest. In today's heavily interconnected world where data are, often by default, entrusted to an online third party, the last bastion of data confidentiality is local storage, as the physical disconnection greatly helps in reducing the room for abusive access. Even there, however, some protection measures need to be implemented, to guard against adversaries who might close that gap. The most trivial example of such an adversary is a thief stealing a user's personal hard disk and reading its raw contents; this is a very simple and well-studied threat model, for which many robust disk encryption solutions exist.

However, disk encryption alone is not enough to handle adversaries empowered by repressive laws or other, less legal methods (e.g., ``rubber-hose''). Unlike the previous scenario, these adversaries gain more than simple ``offline'' access to the disk: they are in a position of power, which they can use to directly and aggressively confront the user about the contents of the protected storage, and by means of (physical, legal, psychological) coercion, they can obtain the encryption keys to any encrypted content identifiable on the user's device. The security goal in this scenario, then, becomes to still retain secrecy of some selected, ``crucial'' data on the disk, by making the presence of such data not even identifiable, thus allowing the user to make up a credible lie about the storage contents. This is exactly the aim of \emph{plausible deniability (PD)}, a powerful security property, enabling users to hide the existence of sensitive information on a system under inspection by overreaching or coercive adversaries.

In the context of secure storage, PD refers to the ability of a user to plausibly deny the existence of certain data stored on a device even under interrogation or forensic inspection of the physical device. The underlying idea is that, if the adversary cannot conclude anything about the existence of hidden sensitive data, they have no motivation to further continue the coercion, thereby (hopefully) limiting the damage for the user. PD was first proposed in 1998~\cite{stego1998} and, since then, many different PD solutions have emerged, attempting to balance the security-efficiency trade-off. One of the most popular PD solutions was TrueCrypt~\cite{truecrypt}, first released in 2004, discontinued in 2014, and replaced by its backward-compatible and technically similar successor VeraCrypt~\cite{veracrypt}.

TrueCrypt and VeraCrypt remain the most well-known PD disk encryption tools available, probably because of the large open source community around them and the good performance, but they suffer from many drawbacks that have been left unaddressed for many years now, both in terms of security and operational model, such as: the possibility of only having \emph{one} layer of extra secrecy, limited filesystem support, and limited functionalities on Linux. In this work we present Shufflecake, a novel disk PD tool that aims at solving these and other limitations by still achieving a good security-performance tradeoff.

\subsection{Motivation}

Albeit less extensively covered by the existing literature, coercive adversarial attacks unfortunately represent many diverse real-world situations. Most of the demonstrable facts, in this regard, concern national security provisions in countries like the USA, France, or the UK, that allow prosecutors to legally oblige citizens to disclose the passwords to their encrypted storage devices~\cite{wiki.ripa, wiki.key_disc_law, wiki.rubber_hose, wiki.in_re_boucher, wiki.us_v_fricosu}, under threats of harsh legal or economic penalties for non-compliance. These provisions are reported to have been often misused, sometimes to the point where people's rightful privacy has been arguably trampled on~\cite{bbc.campaigners, bbc.oliverdrage, bbc.rabbani, guardian.ripa}.

The relative abundance of such reports coming from Western countries, however, is likely due to the comparatively attentive and critical public oversight on the government's operations. In countries with a less-developed system of checks and balances, the precise extent to which the sensitive data of activists, journalists, dissidents, and oppressed minorities are violated, can only be hinted at by the few sporadic cases that occasionally make it to the international headlines~\cite{guardian.msando,refugee}. These kinds of coercive attacks are not merely ``tinfoil-hat paranoia'', but a real-world concern.

\subsection{Previous Work}\label{sec:previouswork}

Different approaches to PD on storage have been proposed, starting from the layer one chooses to intervene at. Digital storage, in fact, is composed of many stacked layers, from the topmost ``logical'' layer (the filesystem) down to the more physical one. Such a layered structure complicates the security analysis, as different PD solutions focus on different layers, each of them leading to different approaches with pros and cons. Certain solutions work at the filesystem layer~\cite{stegfs, defy}; they have to implement a rich interface, made of complex file- and directory-oriented methods (\texttt{fileOpen}, \texttt{fileRead}, \texttt{mkDir}...). Other schemes choose instead to go at the lowest level and modify the FTL~\cite{deftl} (flash-translation layer, for SSDs), but this approach clearly leads to highly vendor-specific solutions. Security of solutions designed for a specific layer might be defeated by adversaries with access to lower layers.

A versatile approach for a robust PD solution is arguably to operate at the \emph{block layer}, whereby the scheme exposes a \emph{block device} interface (a common abstraction layer used by many operating systems to represent storage devices as arrays of fixed-size data \emph{blocks}), providing just a \bRead and a \bWrite method. This is the approach used by solutions like TrueCrypt, and also by Shufflecake; in the remainder of this work, we will only focus on block-layer solutions. In this framework, a single underlying \emph{disk} or \emph{device} is formatted to host one or more \emph{volumes} (logical storage units, usually represented as virtual block devices), each encrypted with a different (usually password-derived) symmetric key. When confronted with the adversary, the user surrenders the passwords to some of these volumes, called \emph{decoy} volumes (because in some use cases they might contain deceptively innocent data): PD security guarantees that, even after these passwords have been given up, inspection of the disk by the adversary still yields no clue whatsoever hinting at the presence of some further, \emph{hidden} volumes. The intuition behind the formal definition of these guarantees, is that the adversary cannot distinguish between the case where all the passwords have been surrendered, and the case where there are still undisclosed ones.

\paragraph{Security models.} Early PD literature has focused mostly on \emph{single-snapshot adversaries}, who are assumed to only check the storage device once. This was considered a natural assumption: in the typical scenario, an activist or journalist is stopped at a checkpoint or arrested and interrogated one time, and her electronics confiscated and analyzed. Provided she manages to escape, she will be on high alert for future investigation, and in particular she will \emph{refresh} the PD protection in use through some specific procedure (e.g., by reformatting the hard disk, or by buying a new device). So, in case of a second check, the adversary will de facto face a completely new \emph{instance} of PD storage, therefore falling back to the single-snapshot case. This is the threat model addressed by solutions like TrueCrypt and VeraCrypt.

The safety of the single snapshot threat model, however, has been questioned in the literature over time, not only because it relies on good user security practice, but also because the technological evolution of storage brought a new issue on the table. In modern devices, especially solid-state disks (SSDs), overwriting a logical sector often results in the underlying physical sector being simply marked as ``unused'' rather than being really overwritten, thereby leaving ``traces'', or ``snapshots'' of the data content at previous points in time. This in turn can (in theory) allow the adversary to break plausible deniability even with a \emph{single} inspection, because by analysing these traces that are left on the device one can see that content at certain locations has changed; since empty, unused space should not change over time, the presence of hidden information therein can be betrayed. This is the scenario considered in \emph{multi-snapshot security} models.

One could argue that multi-snapshot attacks are likely to be very complex, to the point that 100\% evidence of the presence of a hidden volume based only on past sector traces is unlikely to be reached, and an accusation in this sense might not stand in court. 
In fact, we are not aware of a \emph{single} case in public literature of a conviction due to multi-snapshot attacks. On the contrary, there are many documented cases~\cite{tcsatyagraha,tcmiranda,tcjohndoe} where even a simple system such as TrueCrypt was enough to grant acquittal of a suspect. 
This is not to say that single- and multi-snapshot security are equivalent: the latter is stronger. However, one should question what price it is reasonable to pay (in terms of performance, etc) to achieve this stronger security.

\paragraph{ORAMs.} Multi-snapshot attacks are a well-known issue in PD systems (TrueCrypt and derivatives are also vulnerable in this sense) and designing countermeasures turns out to be challenging. As of today, only a few constructions~\cite{sok} achieve multi-snapshot security, but at a hefty performance cost that makes them not practical for most use cases. Most of these solutions are based on \emph{oblivious random access machines (ORAMs)}. ORAMs are cryptographic schemes that aim at hiding the access patterns (in addition to the data content itself) of a trusted agent accessing an untrusted storage. The connection between ORAMs and PD has been investigated since 2014 with the HiVE construction~\cite{hive}. In a nutshell, the idea is that if we use an ORAM to access a device, then nobody, not even a run-time backdoor in the device firmware, can know which (logical) location we access and how, thereby providing a solid method for implementing PD. However, ORAMs are extremely slow: it is known~\cite{yesoramlowerbound} that the bandwidth overhead of any secure ORAM of size $n$ is $\Omega(\log(n))$. The HiVE paper circumvented this problem with the following observation: If we are not worried by run-time backdoors in the device firmware, but are only concerned about ``traditional'' multi-snapshot adversaries, i.e. post-arrest investigation of the device physical layer, then we do not need a fully-fledged ORAM, because \nread operations do not change the state of the device. So all we need is a \emph{``write-only'' ORAM (woORAM)} that only obfuscates \nwrite requests. The advantage is that there is no known efficiency bound for woORAMs, and in fact existing woORAM constructions seem to be slightly better than full ORAMs. The woORAM approach sparked a whole new line of research in multi-snapshot resistant PD solutions~\cite{detwoORAM,sqoram}, and it has been proven~\cite{sok} that resistance to multi-snapshot security under certain assumptions is \emph{equivalent} to the use of woORAMs.

\subsection{Limitations of Existing Solutions}\label{sec:limitations}

So far, the landscape of available PD solutions presents many gaps, both in usability and in security, a fact also hinted at by the relatively scarce adoption of such solutions. By far the most widespread today is VeraCrypt, which comes with many limitations. woORAM-based techniques have been studied in the last few years as promising alternatives to address TrueCrypt and VeraCrypt's security issues. However, it is important to stress that even the most performant woORAM-based schemes are still very slow or wasteful. To put this in perspective: compared to a baseline, HiVE has a slowdown of roughly 200x I/O throughput and wastes 50\% of the disk space, while some recent constructions such as DetWoORAM~\cite{detwoORAM} reach a slowdown of ``only'' 5x but at the cost of wasting 75\% of the disk space. 
This leaves us with the dilemma of either choosing single-snapshot, efficient solutions with limited security, or woORAM solutions with unacceptable performance loss and possibly stronger security.

But woORAMs solutions themselves might not be bulletproof. In fact, we believe that the idea that \nread requests do not change the underlying state of the physical device is a somewhat strong assumption, and hard to justify with modern, complex SSDs that might, for example, cache \nread requests in some undocumented memory area of the firmware, or store read data on an ad-hoc buffer to improve performance through closed-source optimizations, etc.

Another big problem of many plausible deniability solutions (including TrueCrypt) is that the OS itelf (or other applications installed therein) can unintentionally leak to an adversary the presence of hidden data when a hidden volume is unlocked. This can happen for example through the OS logging disk events, search agents indexing files within the hidden volume when this is unlocked, even applications such as image galleries or document readers caching previews of opened documents. Customizing the OS' behavior in such a way to avoid these pitfalls is an almost hopeless task~\cite{schneiertc}. A proposed solution to this problem is to have the OS itself \emph{inside} a hidden volume, which is the idea that led to the concept of ``hidden OS'' on TrueCrypt. However, as far as we know, TrueCrypt (and VeraCrypt) remain the only implementation of this idea, limited to the Windows OS. Overall, we can say that a versatile PD solution able to balance security and usability has been sorely missing for years, especially for Linux, where no really practical solution exists.

\subsection{Our Contribution}

In this work we present \emph{Shufflecake}, a novel PD scheme that aims at striking a balance between the efficiency of TrueCrypt and the security of woORAM-based solutions. Shufflecake operates at the block device layer, like TrueCrypt, but with important improvements:

\begin{enumerate}
\item It offers a virtually unlimited number of hidden volumes per-device, and such number is always hidden to the adversary.
\item Volumes are arranged hierarchically: for the user it is sufficient to unlock one of them, and all the ``less secure'' ones will be unlocked automatically.
\item Unlike TrueCrypt, Shufflecake is \emph{filesystem-agnostic}, meaning that all its features are available regardless of the filesystem chosen by the user.
\item Unlike TrueCrypt, all volumes can be read and written \emph{concurrently} without integrity pitfalls or security degradation.
\item It works natively on Linux, and can be integrated with the kernel for use at boot time and for \emph{hidden operating systems}.
\item Unlike woORAM-based solutions, Shufflecake is extremely fast (with only a minor slowdown compared to a bare, non-PD system) and wastes less than 1\% of the disk space.
\end{enumerate}
This document describes the `Legacy' version of Shufflecake, first introduced in 2022. This version of Shufflecake not only achieves (provable) single-snapshot security, but also implements features that could make possible in the future to achieve a form of ``operational'' (i.e., weak) multi-snapshot security. These features are Shufflecake's hierarchical design and atomic block-rerandomisation, which are not available in tools such as TrueCrypt. We discuss this in~\expref{Section}{sec:conclusion}.

We implemented Shufflecake~\cite{sflcwebsite} in the C language, and we released it as a free software under the GNU General Public License v2+.

\subsection{Acknowledgements}\label{sec:acks}

We are grateful to Edouard Bugnion from EPFL for support and insightful discussions on the Shufflecake scheme, and in particular on the topic of crash consistency. We are also grateful to Vero Estrada-Gali\~{n}anes from EPFL for insightful discussions on the topic of volume corruption. Part of this work was done by E.A. in the context of an EPFL M.Sc. thesis work in the Research Team of Kudelski Security, under official supervision of Edouard Bugnion and technical supervision of T.G..


\section{Preliminaries}\label{sec:preliminaries}

In this section we give the required preliminaries that are going to be used in the rest of this work. In the following, we use ``iff'' as ``if and only if''. 
Array and sequence indices start from $0$. 
By (efficient) {\em algorithm} or {\em procedure} we mean a uniform family of circuits of depth and width polynomial in the index of the family. 
We implicitly assume that all algorithms take the index of the family as a first input, so we will often omit this. 
In the case of cryptographic algorithms, we call such index a {\em security parameter}, and we denote it by \secpar. 
We will often label algorithms with names that reflect their role, e.g. ``adversary'', ``distinguisher'', etc. 
If an algorithm $A$ is deterministic, we denote its output $y$ on input $x$ as $y := A(x)$, while if it is randomized we use $y \from A(x)$; when derandomising an algorithm we look at the deterministic algorithm obtained when considering explicitly the internal randomness $r$ as additional auxiliary input, and we write $y := A(x;r)$. 
We will call {\em negligible} (and denote by $\negl(x)$) a function that grows more slowly than any inverse polynomial in $x$, and {\em overwhelming} a function which is $1$ minus a negligible function. 
Given an event $E$, we denote by $\bar{E}$ its negation. 
Finally, we will write $x \fromunif X$ if $x$ is sampled uniformly at random from a set $X$.

\subsection{Cryptographic Primitives}\label{sec:prelimcrypto}

We assume familiarity with elementary cryptographic constructions and we refer the reader to, e.g.,~\cite{katzlindell} for a more in-depth dive. Here we just recap informally the most relevant concepts.

\subsubsection{Cryptographic security.} We will often define the security of a cryptographic scheme $\Pi$ in terms of a \emph{game}, or \emph{experiment}, that captures the `difficulty in breaking the scheme', leading to so-called \emph{game-based security}\footnote{Other frameworks exist, such as simulation-based, but as a first approximation game-based security notions are very convenient for their intuitivity and simplicity.}. This usually entails comparing the probability of the adversary $\adver$ (modeled as an efficient algorithm) in winning a $\mathsf{Game}$, that is, breaking the scheme, versus the baseline probability of `winning by pure chance', for example by guessing randomly. We call this difference of probabilities the \emph{advantage} of the adversary winning the game for $\Pi$, and we define (computational) security by requiring that this advantage is negligible in $\secpar$ for any (computationally bounded) adversary.
\begin{align*}
\advantage{\Pi}{\mathsf{Game}}{\adver} := \left| 
	\Pr{\adver\left(
		\Pi, \mathsf{Game}
	\right) \to \win} - 
	\Pr{\mathsf{Guess}\left(
		\Pi, \mathsf{Game}
	\right) \to \win}
\right| \leq \negl.
\end{align*}

\subsubsection{Hash functions and KDFs.} A \emph{hash function} is an algorithm that maps strings of arbitrary length into strings of fixed length (e.g., 256 or 512 bits). The most relevant security property for hash functions in our case is \emph{collision resistance}, meaning that it is computationally difficult to find two distinct input strings that map to the same hash value. In order to add resistance against \emph{pre-computation attacks}~\cite{oechslin}, most implementations of hash functions use an additional parameter, called \emph{salt} (usually a non-secret, per-application string of 96-256 bits), to further randomize their mapping. Typical hash algorithms for cryptographic use are SHA256~\cite{sha256}, SHA-3~\cite{sha3}, and BLAKE2~\cite{blake2}.

Hash functions are designed to be very fast and efficient in term of required computational resources. This might actually be an undesirable property when using the function to store images of user-chosen passwords, because it allows for faster adversarial brute-force. In these cases, a \emph{key derivation function (KDF)}, should be used instead. KDFs are functionally similar to hash functions, but are designed in such a way to be \emph{uniformly expensive to compute} on a broad range of computing devices, for example by requiring not only many CPU cycles but also large amount of memory and high latency. Typical KDFs for cryptographic use are Argon2id~\cite{argon2} and Scrypt~\cite{scrypt}.

\subsubsection{Symmetric-key encryption and authentication.} Regarding encryption, the most fundamental primitive is \emph{symmetric-key encryption (SKE)}, also called \emph{secret-key encryption}. An SKE scheme is a pair of algorithms (one for encryption and one for decryption) which define a bijection between a domain (\emph{plaintext space}) and a co-domain (a subset of the \emph{ciphertext space}) of strings. An additional input, the \emph{secret key}, fixes the bijection across the set of all possible ones, and correctness of the SKE ensures that, if the same secret key is used, then the bijection offered by the decryption algorithm is the inverse of that offered by encryption (with overwhelming probability in the case of non-deterministic decryption). The size of a typical secret key in many real-world applications is 128 or 256 bit. This allows to index at most $2^{128}$ or $2^{256}$ unique bijections, which is generally much smaller than the number of possible bijections as the domain space gets larger. For this reason, it is generally impossible to ask the coverage of all possible bijections as a security property of SKEs. Instead, security is usually given in terms of \emph{indistinguishability games}, with strength of the resulting notions depending on the additional power granted to the adversary. One of the most common security notions for SKEs is \emph{indistinguishability under chosen plaintext attack} (or \INDCPA in short). In such game, the adversary's goal is to distinguish between the encryption of two messages of her choice, given additional access to an encryption oracle (for the same, unknown secret key used in the game). The scheme is called \INDCPA secure iff no efficient adversary \adver can successfully win with probability more than negligibly better than guessing at random. Notice that this, in particular, requires that the SKE must be \emph{randomised}, i.e., encrypting twice the same plaintext with the same key will generally yield two different ciphertexts. We will denote encryption (resp., decryption) of a plaintext $\ptxt$ (resp., ciphertext $\ctxt$) with a key $\key$ (and, optionally, a randomness $\rnd$) as $\Encrypt(\ptxt, \key ; \rnd)$ (resp., $\Decrypt(\ctxt, \key ; \rnd)$). The randomness $\rnd$ is generally not needed for decryption, so we will omit it in that case.

If, in addition to privacy, \emph{authenticity} of a message is also required, then SKE is not enough, and \emph{authenticated encryption (AE)} must be used instead. An AE scheme works in a similar way to a SKE, but decryption of a given ciphertext \emph{fails} if the secret key used for decryption is not the same one used to encrypt the original plaintext. When this happens we write that the decryption procedure returns $\bot$. This allows to check that a ciphertext has not been altered or replaced by a malicious adversary, thereby granting authenticity and integrity of the message. A typical way to implement AE is to append a \emph{message authentication code (MAC)} to a ciphertext. A MAC is a random-looking bitstring, for example computed through the \emph{encrypt-then-MAC} procedure with a hash function on a combination of ciphertext and secret key. MACs are useful, among other properties, to check whether a provided key is the correct one to decrypt a ciphertext (without having to actually decrypt the ciphertext first).

\subsubsection{Block ciphers.} As a building block for SKEs, \emph{block ciphers} are widely used. These are algorithms that typically offer two different interfaces, one for encryption and one for decryption. In encryption mode, they take as input a block of plaintext of fixed bitsize $\blocksize$ (the \emph{block size}) and an encryption key of $\keysize$ bits (the \emph{key size}) and return a block of ciphertext, also of size $\blocksize$. This mapping is undone in decryption mode, provided the same key is used. In other words, block ciphers implement a subset of size at most $2^\keysize$ of the space of all possible $(2^\blocksize)!$ permutations (and their inverses) over $\blocksize$-bit strings. One of the most widely used block ciphers are those from the AES family~\cite{aes}, identified by AES-$\keysize$ (with a block size of 128 bits, and a keysize $\keysize$ of 128, 192, or 256 bits).

To turn a block cipher into a generic SKE, a \emph{mode of operation} is required. This is a deterministic procedure that describes how to split input plaintexts or ciphertexts of arbitrary length into fixed size blocks, and iteratively applying the block cipher on these blocks. Typical modes of operation are ECB, CBC, and CTR~\cite{modesofoperation}. Among these, CTR is widepsread for its good characteristics. In order to achieve randomisation as a protection against known-plaintext attacks, many modes of operations also include an additional input called \emph{inizialization vector (IV)}, typically a string of a fixed size, like 64, 96 or 128 bits, not necessarily secret but unpredictable and variable according to the message. Block ciphers can also be used to build AE, but with different modes of operations than the ones used for encryption only, such as GCM~\cite{gcm}. For a given block cipher $\mathcal{B}$ with keysize $\keysize$ and a given mode of operation $\mathcal{M}$, the resulting SKE is usually denoted as $\mathcal{B\mbox{-}M}\mbox{-}\keysize$, for example AES-CTR-192 or AES-GCM-256.

\subsection{Full Disk Encryption}\label{sec:prelimfde}

\emph{Full disk encryption (FDE)} is a security technique that protects the content of a digital storage device (such as a hard drive or SSD) by using encryption `on the fly'. This can include applications, user files, and even the OS itself. The primary purpose of FDE is to prevent unauthorized access to sensitive information in the event of device theft, loss, or unauthorized physical access.

FDE works by employing a cipher (usually a block cipher) to encrypt data at rest on the storage device. A user (or the device manufacturer) must first \emph{initialise} the storage device, by providing an encryption key or passphrase to create and write on the device a special metadata structure that represents an (initially empty) state encrypted with the provided key. In order to protect against space analysis attacks, a very first step before initialisation consists of completely overwriting the device with random noise. Then, every time the system is powered on or the device accessed, the user must provide the valid key or passphrase to decrypt and read/write data on the device. This key is not stored on the device itself and must be entered each time the device is prepared for use (\emph{opening}), usually cached in a volatile and protected area of memory, and thereby erased when the device stops being used (\emph{closing}) or the system is shut down. Except for the one-time initialisation phase (which can be quite slow depending on the device size), the encryption process is typically invisible to the user, as the OS handles the encryption and decryption of data as it is read from or written to the disk. Once the correct key or passphrase is provided, the user can interact with the device normally, without having to manually encrypt or decrypt files, as the OS only exposes a \emph{virtualized} device that looks unencrypted to the user.

FDE can be implemented using hardware-based or software-based solutions. Hardware-based FDE is typically performed by a dedicated encryption chip, while software-based FDE is achieved using encryption software that runs at the operating system level. Implementation is usually done using standard block ciphers contructions like AES.  
Some examples of software-based FDE solutions include BitLocker for Windows, FileVault 2 for macOS, and LUKS for Linux~\cite{bitlocker,luks}. All these implementations have typically only negligible impact on performance compared to a non-FDE system, also thanks to the widespread presence of dedicated CPU instructions to speed up AES computation on personal devices such as smartphones and laptops.

Notice that if the whole OS is protected this way with a software-based solution, then there is a \emph{bootstrapping problem}, because the OS itself cannot natively run while encrypted. This is addressed by either having a small, unencrypted bootloader which launches a minimal FDE application before the rest of the OS can start, or with a lower-level solution usually provided through hardware support such as a \emph{Trusted Platform Module (TPM)}.

\subsubsection{Cipher modes for FDE.} For block ciphers used in disk encryption, the XTS mode of operation \cite{xts} is the most widely adopted because of its performance and security. It avoids the need of explicitly writing IVs for every block on disk by deriving these IVs \emph{pseudodeterministically} from a global IV and sector-dependent metadata. Using CTR mode in a similar way would be a serious security mistake, 
unless care is taken in refreshing (and storing) IVs at every data write, which usually has an impact on performance and space usage. 
The latter approach, however, has a potential advantage: it gives the possibility of \emph{re-randomising} blocks, i.e., changing the ciphertext without changing the underlying plaintext.

\subsubsection{Caveats.} It's important to note that FDE primarily protects data at rest, meaning it is most effective when the device is powered off or in a locked state. It does not provide protection against unauthorized access or data breaches while the system is running and the encryption key or passphrase has been entered. In particular, FDE is arguably less effective on devices such as smartphones, which stay most of the time in an ``on'' state, and offers no protection against malware such as \emph{keyloggers}, which might intercept the password entered by the user, or even access the unencrypted content directly. For comprehensive security, FDE should be combined with other security measures, such as strong authentication, secure boot processes, and proper access control policies.

\subsection{Plausible Deniability}\label{sec:prelimpd}

In this section, we present a formal game-based definition of PD security. It is worth noting that almost every paper in the field has given \emph{its own} security definition, always slightly different from the others; valid attempts have been made to unify them into a single framework~\cite{sok}, but here we will follow the arguably more intuitive one given in~\cite{hive}, which is well suited for the block-layer scenario we work in. In this setting, a user employs a PD scheme to multiplex a single storage device into $\Nvol$ independent volumes $\V_0, \V_1, \ldots \V_{\Nvol-1}$, each $\V_i$ being associated to a different password $P_i$. The PD scheme supports up to $\maxvols$ volumes per device (so $2 \leq \Nvol \leq \maxvols$); the value of $\maxvols$ is publicly known. Both the volumes and the underlying device are block-addressable, meaning that the \nread and \nwrite operations they support have the granularity of a block. The semantics of a scheme $\Pi$ is given as follows.

\begin{definition}[PD Scheme]\label{def:pdscheme} Let $\Nvol \leq \maxvols$, and $P_0, \ldots P_{\Nvol-1}$ user-provided passwords. A \emph{PD scheme} $\Pi$ is a tuple of algorithms:

\begin{itemize}
\item $\Pi.\mathtt{setup}(P_0, \ldots, P_{\Nvol-1}) \to \Sigma$: Initialises disk to host $\Nvol$ volumes $\V_0, \ldots \V_{\Nvol-1}$, encrypted with passwords $P_0, \ldots P_{\Nvol-1}$; returns a device instance description $\Sigma$ which encapsulates everything.
\item $\Pi.\nread(\Sigma, i, \Blog) \to d$: Reads data $d$ in block address $\Blog$ from volume $\V_i$ (we assume $\nread$s to not modify the instance)\footnote{Note this is also the case for woORAM-based constructions, but not necessarily for ORAM-based or other, arbitrary, PD schemes constructions.}.
\item $\Pi.\nwrite(\Sigma, i, \Blog, d) \to \Sigma'$: Writes data $d$ into block address $\Blog$ of volume $\V_i$, and updates the instance.
\end{itemize}
The following correctness requirement applies: for any fixed block $\Blog$ and volume $\V_i$, if $\, \Pi.\nwrite(\Sigma, i, \Blog, d)$ is the most recent write query which precedes a query $\Pi.\nread(\Sigma', i, \Blog)$, then $\Pi.\nread(\Sigma', i, \Blog) \to d$ (for simplicity, we consider operations atomic).
\end{definition}

\subsubsection{Access patterns.} Let us define an \emph{access} as the tuple $o = (\mathtt{op}, i, \Blog, d)$, with $\mathtt{op} \in \set{\nread, \nwrite}$ (if $\mathtt{op} = \nread$, then $d$ is the return value), and $i$ being the index of the volume targeted by the access. Let us also define an \emph{access pattern} as a (chronologically) ordered sequence of accesses $O = <o_0, \ldots , o_{n-1}>$. An empty access $o = \, \perp$ is also defined, which is simply ignored by the instance $\Sigma$.

\subsubsection{PD security}\label{sec:prelimsecpd} The security game for PD inherits some high-level concepts from the \INDCPA game (\emph{ciphertext indistinguishability under chosen-plaintext attack}) for secret-key encryption. The adversary is a distinguisher, and is challenged with deducing whether she is interacting with a $\Sigma$ encapsulating $\Nvol$ or $\Nvol - 1$ volumes. Also, she is allowed to choose the $\nread$ or $\nwrite$ operations to be executed, to capture the idea that indistinguishability must hold no matter the accesses performed on the volumes, hence including adversarial ones. A secret bit $b$ within the game determines whether $\Sigma_0$ (containing $\Nvol$ volumes) or $\Sigma_1$ (containing $\Nvol - 1$ volumes) is first instantiated in the game and made to interact with the adversary. In both cases, we allow the adversary to choose the first $\Nvol - 1$ passwords\footnote{This represents the most unfavourable situation for the user, as we consider these passwords compromised anyway.}, and her goal is to guess $b$.

\begin{experiment}[PD game, generic]\label{exp:pd} 
For a PD scheme $\Pi$ and an adversary $\adver$ the \emph{plausible deniability experiment} $\mathsf{PD}(\Pi,\adver)$ is defined as follows:
\begin{enumerate}
    \item \adver chooses $\Nvol \in \left\{ 2, \ldots , \maxvols \right\}$ and chooses $\Nvol - 1$ passwords $P_0, \ldots, P_{\Nvol - 2}$.
    \item A secret random bit $b \fromunif \set{0,1}$ is drawn. If $b=0$, then an additional secret high entropy password is sampled $P_{\Nvol-1} \fromunif \set{0,1}^\secpar$, where \secpar is the security parameter\footnote{We abuse notation by representing a password as a binary string, but w.l.o.g. it is equivalent to the case of a user-chosen password with \secpar bits of entropy. We assume that the user will choose a high entropy password at least for the hidden volume.}.
    \item\label{algpoint:instcreate} The game creates $\Nvol - b$ volumes: $\Pi.\mathtt{setup}(P_0, \ldots, P_{\Nvol - 1 - b}) \to \Sigma_b$
    \item \adver performs interactive rounds of queries. Every query works as follows:
    \begin{enumerate}
        \item\label{algpoint:constraints} \adver chooses access patterns $O_0$ and $O_1$, where $O_0$ is, in the adversary's intentions, aimed at $\Sigma_0$ (thus potentially containing some operations on $\V_{\Nvol-1}$) and $O_1$ is aimed at $\Sigma_1$ (and so only contains operations on $\V_0 , \ldots , \V_{\Nvol - 2}$). 
        
        She also chooses a bit $v$, signalling whether she wishes a snapshot of the disk at the end of this round.
        
        These adversarial choices are subject to constraints, which we discuss in the next paragraph.
        \item The game only executes $O_b$ (on $\Sigma_b$, the only instance that was created in step~\ref{algpoint:instcreate}). If requested, it sends the resulting disk snapshot $D$ to \adver.
    \end{enumerate}
    \item At the end of all rounds, the adversary outputs a bit $b'$.
    \item The game outputs $1$ iff $b = b'$, $0$ otherwise.
\end{enumerate}
\end{experiment}

Here we have omitted the constraints that the adversary is subject to in step \ref{algpoint:constraints}, when choosing the access patterns and when choosing the snapshot bit $v$; we will present them now. Without any such constraints, we will see that security would be impossible to achieve; also, the exact set of constraints will modulate the induced threat model.

\paragraph{Constraints on the bit $v$.} This constraint governs the snapshotting capabilities of the adversary, thus the adversary power. We only define two extreme cases:

\begin{enumerate}
    \item Arbitrary - No constraint: the adversary is allowed to set $v=1$ in all of the interactive rounds. This is the strongest form of multi-snapshot security, as the adversary can obtain a snapshot any time she desires. 
    \item One-Time. The adversary is single-snapshot, i.e. can only set $v=1$ for one of the interactive rounds.
\end{enumerate}

\paragraph{Constraints on the access patterns.} These constraints define the adversary goal, by specifying which two exact situations she has to distinguish between: if a PD scheme is secure (i.e., the adversary cannot distinguish) under the game enforcing such a constraint, the implication is that a user, having performed some access pattern $O_0$ including some operations on $\V_{\Nvol-1}$, can plausibly claim to instead have executed a corresponding $O_1$, which only accesses the volumes $\V_0, \ldots , \V_{\Nvol - 2}$ (whose passwords have already been surrendered).

\paragraph{Discussion on the constraints.} Let us first clarify that some constraint is necessary in order to have any hope in the PD game against the adversary. Otherwise, the adversary could submit $O_0$ and $O_1$ containing completely different (logical) $\nwrite$ accesses to the decoy volumes $\V_0, \ldots , \V_{\Nvol - 2}$, and there would be no way of making the two outcomes indistinguishable, since the adversary holds the passwords to those volumes, so it could trivially verify which of the two patterns was executed. This suggests the need for a minimal rule, stating that, whether $O_0$ or $O_1$ is executed, the resulting logical contents of the decoy volumes $\V_0, \ldots , \V_{\Nvol-2}$ must be the same. From the user's perspective, this basic requirement means that we do not try to disguise the accesses 
to decoy volumes as something else, both because we do not need to, and because there would be no way of doing it even if we wanted to.

Furthermore, we notice that many PD solutions (including those based on woORAMs) treat $\nwrite$ requests in a completely different way than $\nread$ requests, i.e. a $\nwrite$ request could \emph{trigger allocation or reshuffling} of a certain volume sector. This might happen without breaking the minimal rule described above: if an adversary first $\nread$s at a previously unallocated position $\Blog$, obtaining data $d$, and then $\nwrite$s the same data $d$ at the same location $\Blog$, this might cause a change of state in the instance without changing its logical content (as mandated by the minimal rule above). This, in turn, would enable an attack where the adversary merely checks whether this state change happened or not. However, we must consider this attack trivial: detecting this kind of state change on a decoy volume would actually not compromise security. Therefore, in order to capture this concept, we also demand that the set of all blocks that are touched by $\nwrite$ requests be the same for both $O_0$ and $O_1$, for all volumes $\V_0 , \ldots \V_{\Nvol - 2}$. We stress that this extra constraint is also very minimal and does not weaken the security guarantees offered in practice; in fact, many previous security definitions 
demand even stricter constraints, for example by demanding that $O_0$ and $O_1$ are of the same length, which we do not require here.

\subsubsection{Single-snapshot security}\label{sec:ss}
In the single-snapshot case, the security game can be simplified. More formally, if the ``One-Time'' constraint is enforced on bit $v$, then \expref{Experiment}{exp:pd} is equivalent to the same game where the number of interactive rounds is set to 1 instead of being chosen by the adversary. This is because the access patterns submitted \emph{before} obtaining the (only) disk snapshot might as well be all concatenated into one, as they cannot be chosen adaptively; moreover, all the access patterns submitted \emph{after} obtaining the snapshot can be disregarded altogether, as their effect will not be detectable by the adversary.

Security in this one-round game can then be rephrased as follows:

\begin{dfn}[Single-Snapshot Security]\label{def:ss} A PD scheme $\Pi$ is \emph{single-snapshot (SS)-secure} iff for any PPT adversary \adver which chooses $\Nvol \in \left\{2,\ldots,\maxvols\right\}$, passwords $P_0, \ldots , P_{\Nvol-2}$, and access patterns $O_0$ and $O_1$ subject to the constraints outlined in \expref{Section}{sec:prelimsecpd}, it holds:
\begin{align}
	\left| \Pr{\adver\left(D_0\right) \to 1 } - \Pr{\adver\left(D_1\right) \to 1 }\right| \leq \negl,\label{eqn:1}
\end{align}
where:
\begin{itemize}
\item $D_0$ is the disk snapshot resulting from the application of $O_0$ to $\Sigma_0$, where $\Sigma_0 \from \Pi.\mathtt{setup}(P_0, \ldots, P_{\Nvol-1})$ and $P_{\Nvol-1} \fromunif \{0,1\}^\secpar$.
\item $D_1$ is the disk snapshot resulting from the application of $O_1$ to $\Sigma_1$, where $\Sigma_1 \from \Pi.\mathtt{setup}(P_0, \ldots, P_{\Nvol-2})$.
\end{itemize}
\end{dfn}

\section{TrueCrypt and VeraCrypt}\label{sec:tc}

Given its relevance in the context of comparison to Shufflecake, we want to discuss here TrueCrypt~\cite{truecrypt}, which was the first disk encryption software (now discontinued) to offer PD capabilities. It was developed around the early 2000s, before BitLocker~\cite{bitlocker} and LUKS~\cite{luks} became the default standards for disk encryption on Windows and Linux, respectively. Its development has come to a sudden halt in 2014, but a backward-compatible successor exists (VeraCrypt~\cite{veracrypt}) that has kept most of the design principles, and improved on some minor aspects (like a stronger key derivation). For our purpose in this work, we will focus on TrueCrypt only, as all our considerations similarly apply to VeraCrypt.

TrueCrypt can operate in two main modes: with ``standard'' (sometimes called ``outer'') encrypted volumes, or with ``hidden'' volumes. In the former case, it is functionally similar to other FDE solutions like LUKS (but with random-looking encrypted headers). In the latter case, a hidden volume is embedded in the unused empty space left by the content of the decoy standard volume. Plausible deniability is given by the fact that disk headers and content are indistinguishable from random, which makes it hard to distinguish between the two cases without the correct passwords.

\subsection{Design}\label{sec:tcdesign}

TrueCrypt (like Shufflecake and many other existing PD schemes) works as a \emph{stacking driver}, that is, a device driver operating on top of another device driver. It exposes a \emph{logical} (virtual) storage space to the upper layer, which directs \emph{logical} \texttt{TcRead} and \texttt{TcWrite} requests to it; the stacking driver then executes its algorithm to map these requests to \emph{physical} block \texttt{bRead} and \texttt{bWrite} requests to the underlying device driver, which manages the \emph{physical} storage space. Here the distinction between \emph{logical} and \emph{physical} is the distinction between before and after the translation operated by the stacking driver, regardless of whether the \emph{physical} storage space is also a virtual device.

The first initialisation operation performed by TrueCrypt when creating new volumes within a device is to fill the disk with random bytes, which is also the case for regular disk encryption tools including LUKS, as we already discussed. The first part of the disk contains the fixed-size encrypted header of the standard volume, and an equal-size empty slot filled with random bytes (remaining from the initialisation procedure). Then comes the actual encrypted data section of the standard volume, which includes some empty space, also filled with random bytes (coming from the initialisation procedure).

\begin{figure}[ht]
\centering
\includegraphics[width=0.65\textwidth]{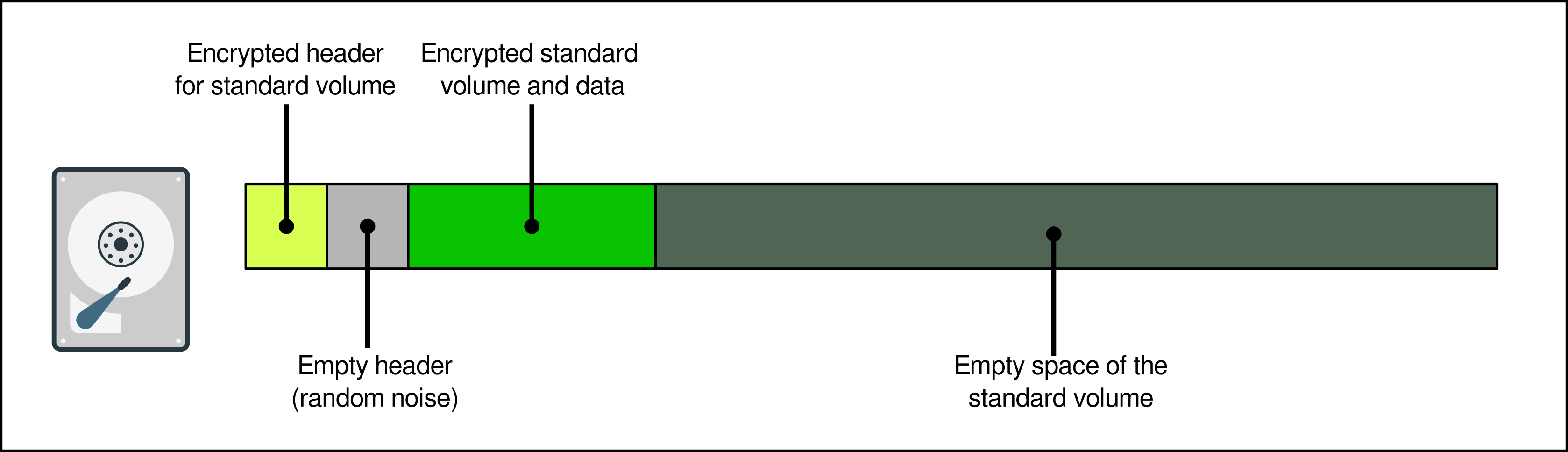}
\caption{TrueCrypt's disk layout (standard volume).}
\label{img:tc1}
\end{figure}

TrueCrypt optionally allows to ``embed'' a hidden volume in the (contiguous) empty space left by the standard volume: this is the mechanism providing plausible deniability. Its encrypted header then fits in the empty slot left after the header of the standard volume. The standard volume and the hidden volume are encrypted with two different passwords.

\begin{figure}[ht]
\centering
\includegraphics[width=0.65\textwidth]{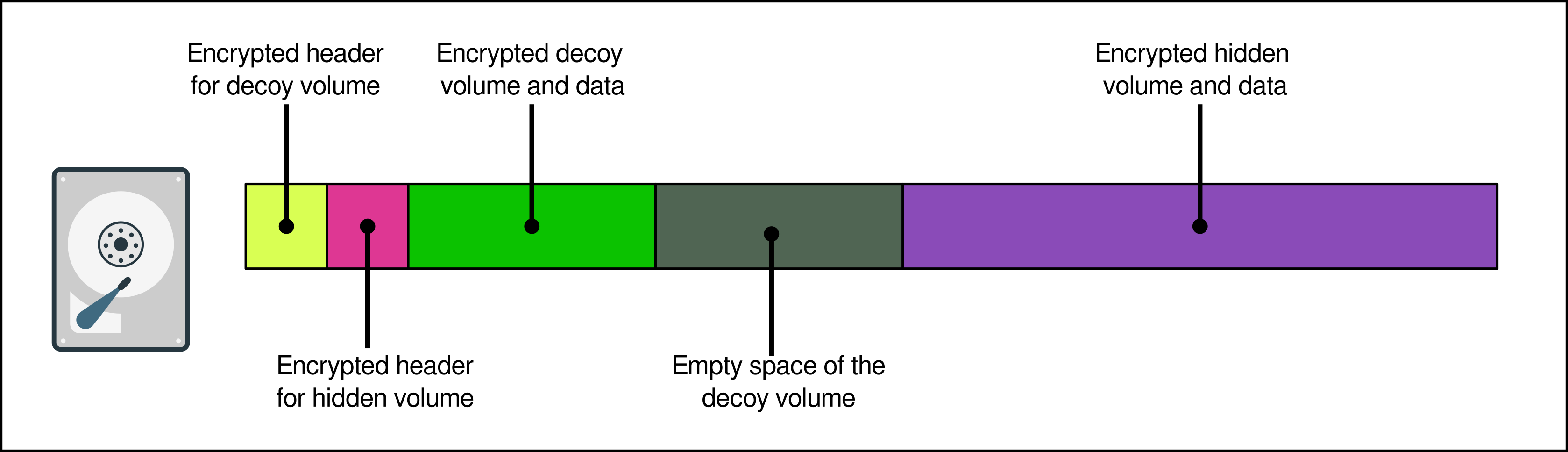}
\caption{TrueCrypt's disk layout (standard/decoy and hidden volumes).}
\label{img:tc2}
\end{figure}

One big limitation of TrueCrypt is that it can only support a hidden volume if the outer volume is formatted as a FAT filesystem\footnote{Latest versions of TrueCrypt on Windows also support NTFS-formatted outer volumes, albeit with serious limitations, among which a waste of 50\% volume space.}. This is because we need the empty space left by the decoy volume to be contiguous. Most modern filesystems (like ext4) ``jump back and forth'' on the disk as they are written with data, leaving lots of empty blocks, or ``holes'', in the middle. Instead, the FAT filesystem is special, in that it grows incrementally and, up to the physiological holes created by deleted files, occupies all the space up until its last utilised block. This way, one can have the hidden volume start at a certain offset, after the end of the decoy volume, and then follow data allocation linearly. TrueCrypt automatically computes a convenient starting position for the hidden volume (leaving some leeway buffer after the end of the standard volume), and places it among the metadata of the hidden volume's header. Whereas the hidden volume is assigned a logical size that follows the physical space actually allocated to it, the standard volume is \emph{not} resized, and keeps logically mapping onto the whole disk. This is crucial in order to not defy deniability: if we resized the standard volume, this information (which leaks the existence of a hidden volume) would be written in the metadata of its file system, which is inspected by the adversary.

Another big limitation of this approach is that, from the moment that a hidden volume is embedded into a standard volume, the latter will be very limited in the possibility of content growth: as the hidden volume lives in the empty space of the standard volume, this seemingly empty space can never shrink (except for the leeway buffer), or the hidden volume will be corrupted.

\subsection{Operational Model}

Using TrueCrypt comes with restrictions on what the user can do in order to preserve data integrity and plausible deniability.

As already mentioned, once the hidden volume is created, its starting position is final, and it ``freezes'' the end of the decoy standard volume, limiting the maximum size that it will ever be able to attain. Since the standard volume cannot be resized to accommodate for the hidden volume, and instead keeps mapping onto the whole disk, it is up to the user to not let it grow too much and overwrite the hidden volume. This is achieved both by frequent de-fragmentation of the standard volume, and by actually not writing too much data into it. If one wants to be absolutely safe about data corruption within the hidden volume, the recommendation for the user is to never unlock only the standard volume (except if under coercion), but to always unlock either 1) the hidden volume only, or 2) both hidden and standard volumes, but keeping the latter read-only.

The standard volume must contain ``decoy'' data, that will reasonably convince the adversary that it is the only volume existing on the disk. Clearly, the user should only surrender the password of the standard volume to the adversary when under coercion. This, in turn, opens the door for corruption of the hidden volume. Eliminating completely such corruption risk is unavoidable, in TrueCrypt this can can only be mitigated by frequent backups.

\subsection{Security}

Here, we analyse TrueCrypt's security under different threat models.

\subsubsection{Pseudorandomness.} Unlike other FDE solutions such as LUKS and Bitlocker, TrueCrypt-formatted devices do not contain any cleartext header. This means that a TrueCrypt-formatted device is indistinguishable, when at rest, from a device completely filled with random noise. This feature is desirable in certain scenarios, for example it is more straightforward and less risk-prone if one wants to embed a TrueCrypt container file within another medium using steganography, but represents a tradeoff against ease of integration with other parts of the system, which is the approach preferred by all-purpose FDE solutions such as LUKS and Bitlocker. Anyway, it must be stressed that this feature is not relevant per se for PD, because in the PD scenario the adversary is always provided with at least \emph{one} decryption key.

\subsubsection{Single-snapshot security.} In TrueCrypt, once the user surrenders the password of the decoy volume and lets the adversary decrypt it, the only part of the disk contents that remains to be ``interpreted'' by the adversary is the non-decrypted space 
in the unused areas of the decoy 
file system. However, whether this space is actually empty or whether it contains a hidden volume, it will be filled with random bytes that are not readable with the decoy password alone. Therefore, even if a hidden volume is present, the user can \emph{plausibly claim} that the remaining space is empty and filled with random bytes: the adversary has no way to disprove this claim, \emph{or even to question its likelihood}, based on the observed disk content.

\subsubsection{Multi-snapshot security.} It is easy to see why TrueCrypt is insecure in the multi-snapshot threat model: what happens if the adversary obtains two snapshots of the disk at two different points in time, and the user has made changes to the hidden volume in the meantime? By comparing the two snapshots, the adversary clearly sees that some of the allegedly ``empty'' blocks have changed, which immediately reveals that a second volume exists, because TrueCrypt never re-randomises the actually-free space.

\subsubsection{TrueCrypt's hidden OS.} Latest versions of TrueCrypt offer a solution to the OS' tendency to leak the existence of hidden partitions through a ``hidden OS'' feature. A decoy OS is installed within a standard volume, while a separate OS is installed within the hidden volume of another partition. In order to decide which OS is booted according to the provided password, the computer's bootloader is replaced by the ad-hoc \emph{TrueCrypt bootloader}, which will first try to boot the decoy OS with the user-provided password and then, if unsuccessful, will try to boot the hidden one. Since the decoy OS itself never sees the hidden partition, there is no possibility for it to even be aware of the existence of hidden data. Notice, however, that this feature is only available for Windows.

\subsection{Other Limitations}\label{sec:tclimitations}

Here we discuss other problematic aspects of TrueCrypt. It is to be noted that some of such considerations apply today, but were less relevant in the early 2000s, when TrueCrypt was first conceived.

First of all, as discussed, the standard volume must be formatted as a FAT or NTFS filesystem if a hidden volume is desired. However, FAT is now outdated: it used to be widespread, and it still is on cheap electronics and some data transfer media, but today there is little plausible reason to use it on a personal computing device. Therefore, the mere fact of using FAT raises a red flag. The NTFS option is only available on Windows and comes with other serious limitations.

Another problem arises from the fact that the user must avoid or limit the use of the decoy partition in order to not corrupt the hidden one. Yet, decoy volume(s) must ``look legitimate'': it must be plausible by looking at their content that they are the only ones. In particular, they must be reasonably up to date: if we only ever work on the hidden volume, and completely forget about the decoy, an adversary unlocking the decoy would become very suspicious seeing that the most recent updates are months if not years old.

In general, it is not within the scope of PD schemes to hide \emph{themselves}, i.e., hide the very fact that that scheme is being used. We must assume that this fact is known to the adversary, who might, e.g., by searching the user's laptop, discover a TrueCrypt installation. Since a locked TrueCrypt volume is indistinguishable from random data, when asked for the first password, we could in theory even claim that the disk is not formatted with TrueCrypt at all, but is instead the result of a secure wiping procedure, or even that it is a pool of random data coming from other sources. However, a real-world adversary will arguably be unconvinced given the knowledge that a system like TrueCrypt is in use. So, a recommended course of action is to separate the TrueCrypt-supporting system (e.g., a laptop) and the encrypted media (e.g., a USB stick). This might be cumbersome for most use cases.

Another big limitation is the fact that only \emph{one} hidden volume within each standard volume is supported. This is a problem, because the adversary might reasonably suspect that TrueCrypt is in use \emph{exactly} to hide something through its PD feature: if we only meant to encrypt a volume, we would reasonably use a more widespread and supported solution like LUKS or BitLocker. A user could claim that they prefer using independent, niche open-source software to secure their data, but the safest course of action is assuming that the adversary might not believe this claim, and will pressure for a second password.

The short answer to this problem is: a robust TrueCrypt-like PD solution should allow \emph{simultaneous access} to \emph{more} than two layers of volumes, without revealing \emph{how many}. That way, we can create a series of volumes with increasingly ``private'' contents (that could well be all decoys), so as to reveal more than one password to the adversary and convince them, based on the resulting decrypted contents, that we have effectively given up what we were hiding, while in fact we are still holding the password to one more top-secret volume whose existence they have no more reason to suspect.


\section{The `Legacy' Shufflecake Scheme}\label{sec:sflc}

In this section we present (the `Legacy' version of) our Shufflecake scheme, we explain its way of operation, and we provide a security analysis. 
We do not provide concrete parameters in this section, as we want to present a generic framework that can be adapted to various implementations, but we provide our own choice of parameters in~\expref{Section}{sec:implementation}. In particular, in this section we will treat security as asymptotic, without targeting a concrete bit-security level. 
We assume a target storage block device of \devbsize blocks, each of them of size \blocksize bits, and a maximum of \maxvols supported volumes per device.

\subsection{Design}\label{sec:sflcdesign}

By \emph{device}, we mean the underlying disk, which exposes a \emph{physical} storage space. Instead, \emph{volumes} are the \emph{logical} storage units that map onto a device. 
The name `Shufflecake' stems from the analogy of mixing up \emph{slices} of a cake (the device) in order to provide many stacking \emph{layers} of privacy (the volumes). 
Conceptually, Shufflecake's operation consists of four functionalities:

\begin{enumerate}
\item \emph{Initialize} a device: this is done only once, when a new device is first prepared for use in Shufflecake. It consists in overwriting the device with random data, asking the user to provide the number $\Nvol$ of desired volumes and related passwords, and creating an encrypted header with metadata using this information.
\item \emph{Instantiate} a device: this is the preliminary stage of preparing a Shufflecake-initialized device for use. It consists of reading the user's provided password, trying to decrypt the device's header metadata with the derived key, and, if successful, recover information on the available volumes provided.
\item \emph{Open} a volume: using the correctly derived volume key, volume-specific metadata is read from the relevant header section. This metadata is used to create a logical device which is presented to the user, and the user's OS can issue \SflcRead and \SflcWrite requests to this logical volume.
\item \emph{Close} a previously instantiated device: ephemeral state changes, if present, are written (encrypted) to disk, and then \emph{all} the open volumes provided by that device are removed from the user's view.
\end{enumerate}

At its core, Shufflecake is a block indirection layer on top of an encryption layer. Our indirection layer realizes a mechanism which is already a strong improvement over TrueCrypt, since it fixes two of its crucial limitations: it allows simultaneous use of multiple volumes, and it is filesystem-independent. Decryption keys for every volume are derived by a password and other randomness. Furthermore, the decrypted payload of the header of volume $\V_i$ with $i > 0$ also contains a copy of the header decryption key for volume $\V_{i-1}$. This allows to recursively open all volumes present in a device by using a single password, which in turn improves security and user experience, as we will see.

\subsubsection{Disk Layout.}
The device's physical storage space is statically divided into a \emph{header section} and a \emph{data section}. The header section is found at the beginning of the disk, and it is composed of a fixed-size \emph{device master block (DMB)}, and $\maxvols$ equal-sized \emph{volume headers} (each of them comprised of a \emph{volume master block (VMB)} and a \emph{slice map}), irrespective of how many volumes there are effectively. This mild waste of space is necessary in order to to prevent the adversary from trivially deducing the number of volumes by the size of the device header (which might be possible by analysing the data allocation pattern even when not all volumes are opened). Let us analyse all these sections, starting with the data.

\begin{figure}[ht]
\centering
\includegraphics[width=0.8\textwidth]{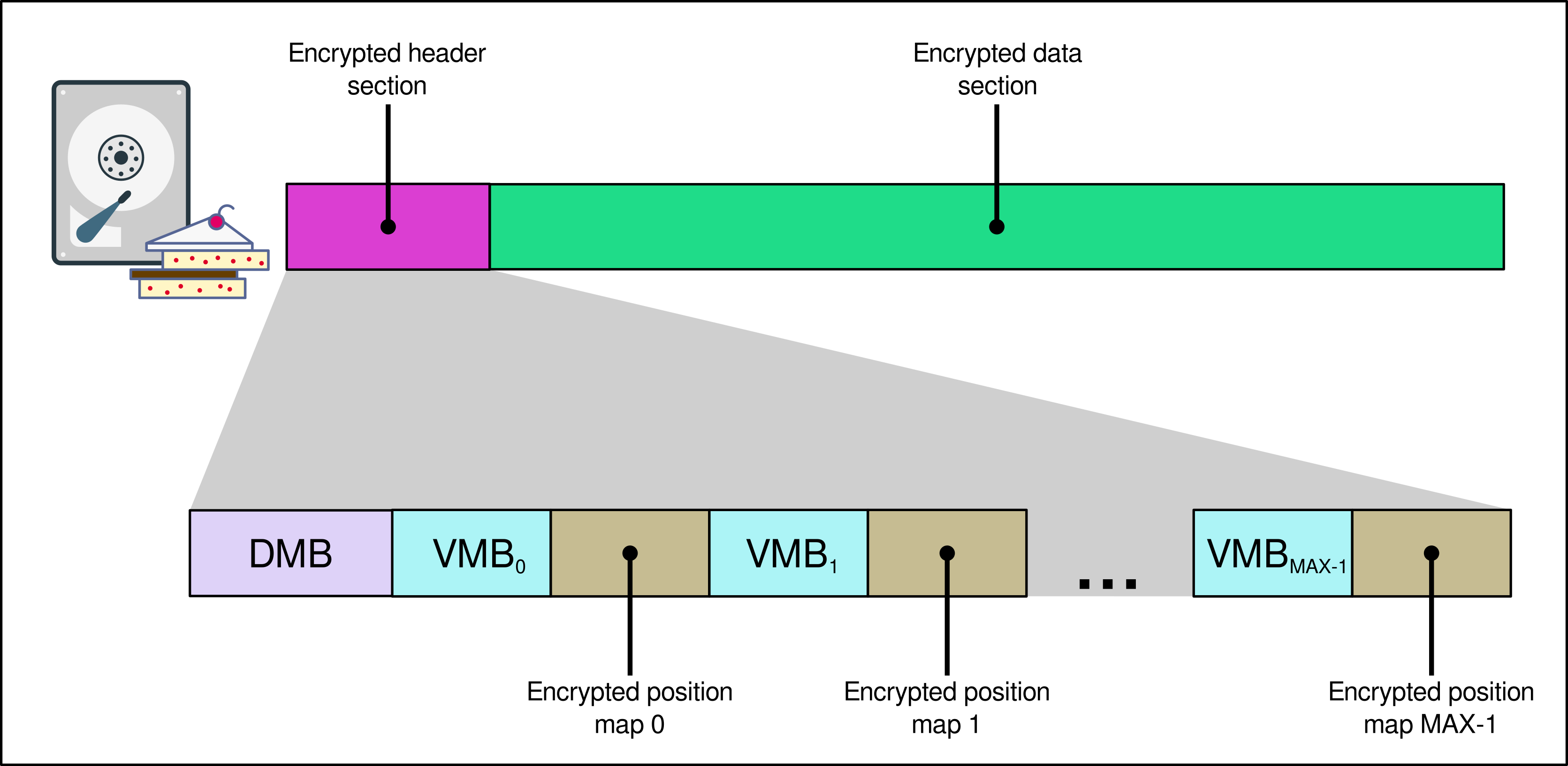}
\caption{(Legacy) Shufflecake's disk layout overview.}
\label{img:sflclayout}
\end{figure}

\paragraph{Data section.} Instead of mandating that the volumes be physically adjacent on-disk, like in TrueCrypt, we randomly interleave them as encrypted (but not authenticated) fixed-size \emph{slices}, where every slice belongs to one volume and contains a certain number of blocks. Metadata in the $i$-th volume header allows to reconstruct the logical content of $\V_i$ by mapping the corresponding slices as a (virtual) contiguous space.

\begin{figure}[ht]
\centering
\includegraphics[width=0.8\textwidth]{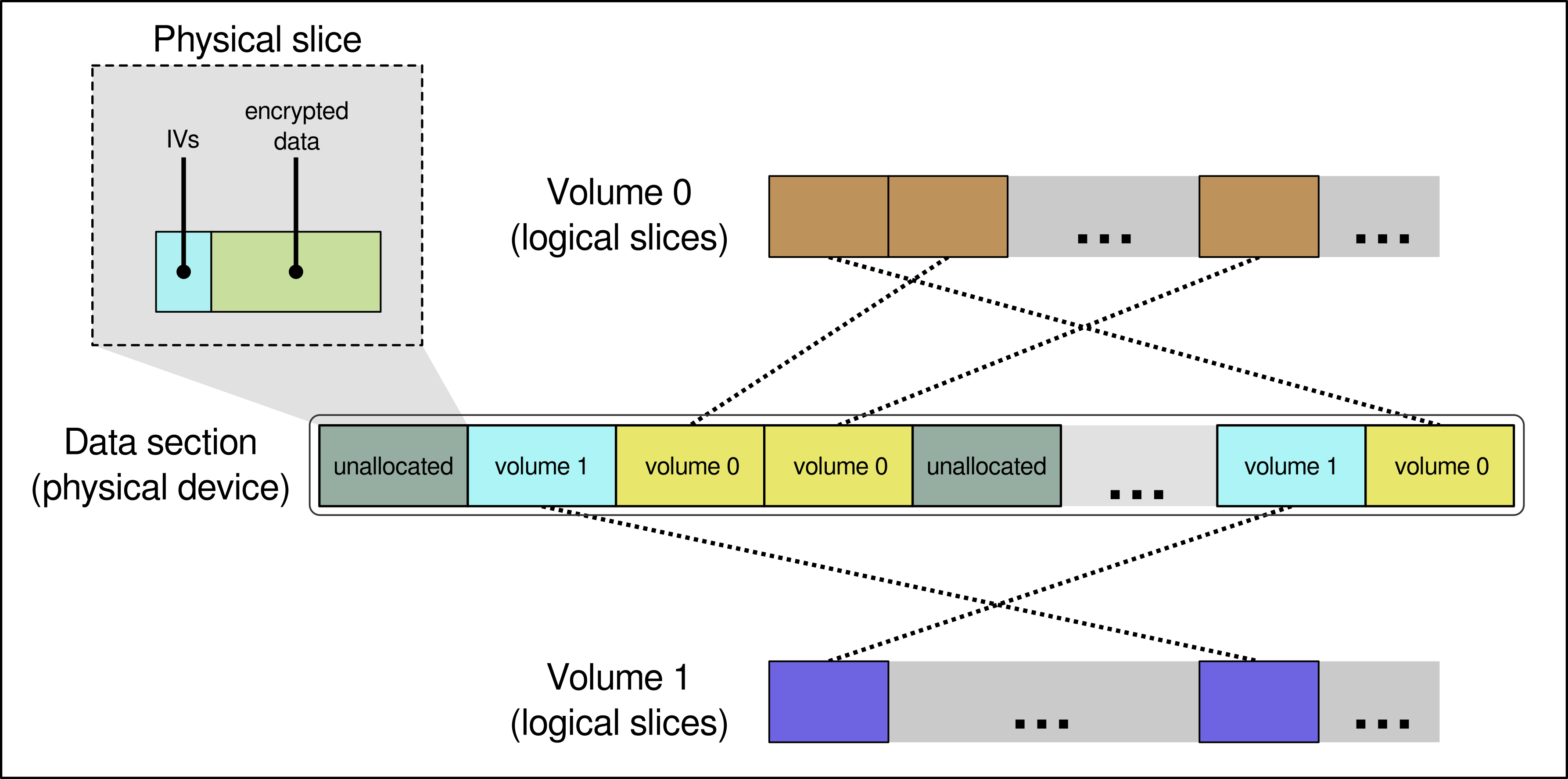}
\caption{(Legacy) Shufflecake's data section layout and slicing.}
\label{img:sflcdata}
\end{figure}

In Shufflecake `Legacy', we distinguish between \emph{logical slices}, of size $\Sl$ blocks, and \emph{physical slices}, of size $\Sp = \Sl + \deltas$, where $\deltas$ blocks are used to store the encryption IVs for that slice. Logical slices store the data of their respective volumes, while physical slices are used to reserve space on the disk to allocate the encrypted logical slice. A physical slice can either be unallocated (i.e., unclaimed by any volume), or mapped to a single logical slice belonging to some volume $V_i$. In the latter case, the mapping from the \emph{logical slice index (LSI)} $\LSI$ of volume $\V_i$ to the device's \emph{physical slice index (PSI)} $\PSI$ is given by a correspondence $\SliceMap_i: \LSI \mapsto \PSI$ which is just a lookup of $\V_i$'s slice map (basically, a per-volume array). I/O operations to a \emph{logical block address} $\Blog$ of volume $\V_i$ are performed through the two interfaces $\SflcRead$ and $\SflcWrite$. We will describe later these two interfaces, as well as the structure of the slice map.

\paragraph{Device master block (DMB)} The DMB encapsulates all password-related data. It begins with one single KDF salt, shared for all volumes: this salt, combined with a volume $\V_i$'s password through a KDF, yields the volume's \emph{key-encryption-key} ($\KEK_i$). Notice that we derive every $\KEK_i$ by using just a single global salt for all $\maxvols$ volumes, otherwise we would incur in up to $\maxvols$ expensive different key derivations every time we instantiate a device\footnote{This limitation can actually be avoided by using some key derivation tricks, like \emph{`re-salting'}, i.e. using the output of the KDF in combination with a (fast) hash using a per-header salt. We leave this approach for future consideration.}. This does not hamper security, 
as password hashes are never stored on disk, but only used to generate the KEK which is in turn used as a decryption key.

Then come $\maxvols$ DMB \emph{cells}, each being an authenticated ciphertext (together with the corresponding IV), encrypted with the respective volume's KEK: the plaintext is itself another cryptographic key, encrypting the volume's VMB (the \emph{volume master key $\VMK_i$}). This key decoupling allows us to possibly change a volume's password without having to re-encrypt all its content with a different key. For granularity and consistency, the overall size of the DMB is fixed to be exactly one block. The rest of the DMB, and the unused DMB cells, contain random noise.

\begin{figure}[ht]
\centering
\includegraphics[width=0.8\textwidth]{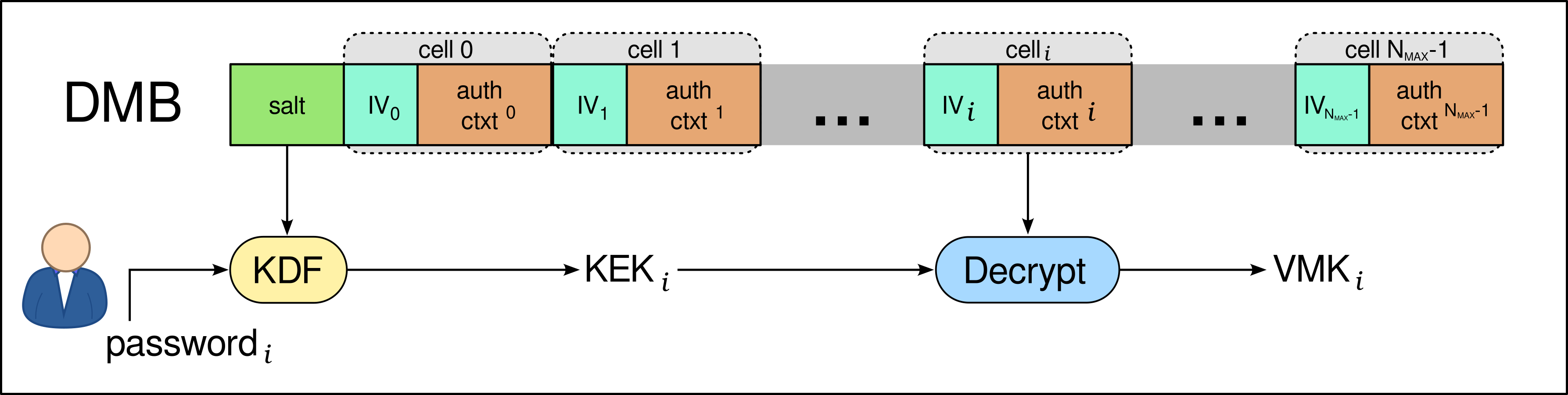}
\caption{(Legacy) Shufflecake's Device Master Block (DMB).}
\label{img:sflcdmb}
\end{figure}

\paragraph{Volume master blocks (VMBs).} The header of volume $\V_i$ is composed of a Volume Master Block (VMB), followed by the encrypted (but not authenticated) slice map of $\V_i$. We discuss the slice map in the next paragraph. The VMB is a single block containing a (non-authenticated) ciphertext, encrypted with the volume master block's key $\VMK_i$, and the associated IV. The underlying plaintext is composed of the following fields:
\begin{itemize}
\item The volume's \emph{volume-encryption key ($\VEK_i$)}, used for encrypting the actual data section and the slice map.
\item The previous volume's VMB key $\VMK_{i-1}$ (or a random value if $i = 0$).
\item The number of slices $\numslices$ contained in the device.
\item The remaining space up to filling the block size is left random, but can be optionally used to embed additional volume-related metadata if needed.
\end{itemize}

The device-specific value $\numslices$ defines and fixes the size of the slice maps, even in the case that the device is resized; it is replicated across all volume headers in order to be decryptable with any provided password. 
The presence of $\VMK_{i-1}$ is what allows us to impose a hierarchy on the otherwise-independent volumes (they are all treated equally in the data section). This way, once we open volume $\V_i$, we can iteratively walk the backwards-linked list induced by this field, and also open volumes $\V_{i-1}$ through $\V_0$. While this approach compromises deniability for a volume $\V_i$ ``in the middle'', once the password to $\V_j \, , \, j>i$ has been provided, it does not harm deniability as defined by the security game: the volume we want to hide is the last one, $\V_{\Nvol-1}$, not the middle ones. The usefulness of this approach is discussed in~\expref{Section}{sec:opmod}.

\begin{figure}[ht]
\centering
\includegraphics[width=0.8\textwidth]{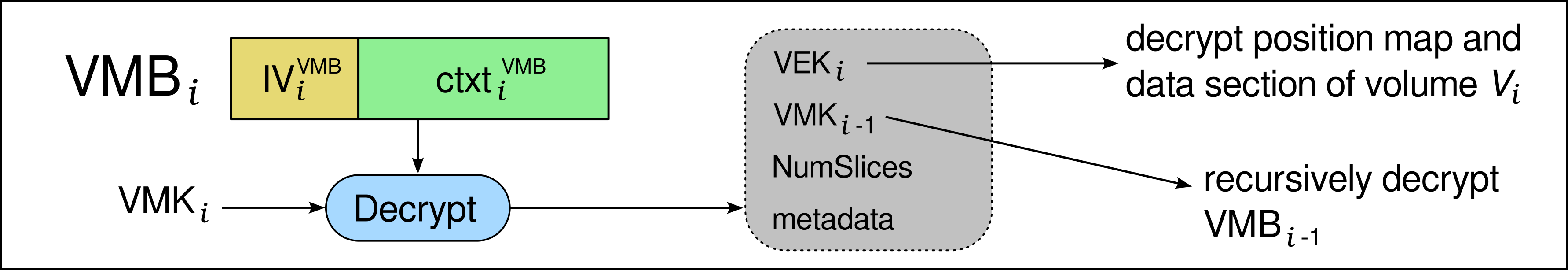}
\caption{(Legacy) Shufflecake's Volume Master Block (VMB).}
\label{img:sflcvmb}
\end{figure}

\vspace{-0.5cm}

\paragraph{Slice maps.} Slice slice maps are arrays $\SliceMap_i$ of $\numslices$ elements, where every element is a PSI: the index of each element is the LSI mapping to that PSI. Each volume's slice map is decrypted and loaded in memory when the volume is \emph{instantiated}; these maps are kept entirely in RAM while the volumes are ``live'', and persisted on-disk (encrypted, together with their fresh IVs) when the volumes are closed. slice maps are stored after the VMBs as equal-size ciphertexts, large enough to address all the possible physical slices. The RAM and disk space requirements are modest: if the underlying device is $\devbsize$ blocks large and the header size is \Shdr blocks, there can be at most $\maxslices = \lceil \frac{\devbsize-\Shdr}{\Sp} \rceil$ physical slices, each requiring $O ( \log \frac{\devbsize}{\Sp} )$ bits to be indexed, for a total size of $O ( \frac{\devbsize}{\Sp} \log \frac{\devbsize}{\Sp} )$ bits per slice map. This is in turn due to the choice of addressing the storage space at the slice granularity, instead of the block granularity, which would have entailed a slice map of $O ( \devbsize \log \devbsize )$ bits.

\subsubsection{Operation.} We now describe more in detail how (`Legacy') Shufflecake operates and explain the rationale behind certain design choices. 
We start by explaining the use encryption on data blocks. Then we look at the indirection layer between physical data on disk and data on the logical volumes, i.e., the mechanism which creates the correspondence between physical and logical slices. We explain how this correspondence is updated when new logical slices are needed as data is written on the volumes. Then, we explain how Shufflecake handles the case of allocation errors, i.e., when a physical slice is mistakenly assigned to more than one volume. Finally, we detail how the instantiation procedure works.

\paragraph{Cryptographic layer.} As in many disk encryption solutions, we encrypt with the block granularity, meaning that blocks are the unit of both I/O requests and encryption/decryption; in other words, one IV encrypts one block. Many disk encryption schemes generate these IVs pseudo-deterministically from some public context information and the volume's secret key (e.g., this is what happens in the XTS mode of operation) in order to save the space and the I/O overhead needed to store and retrieve them. Having such a deterministic procedure for generating IVs on the fly is enough for the threat model covered by FDE and single-snapshot secure PD solutions like TrueCrypt.

In our `Legacy' case, however, we stick to explicitly random IVs because we want to keep the future possibility to extend Shufflecake with some degree of multi-snapshot security, requiring us to re-encrypt some blocks with a different IV while leaving their content unchanged. For this reason, we use CTR mode instead, and the IV of a block is refreshed at each \SflcWrite for that block, to avoid IV-reuse attacks. This means that all these IVs are stored on-disk.

This strategy introduces a potential performance issue: a naive implementation would translate each logical \SflcWrite to a physical \bWrite of the corresponding physical data block, plus an additional \texttt{bRead}-update-\texttt{bWrite} of the \emph{whole} corresponding IV block. This would be very wasteful in terms of I/O overhead, because we only need to update one IV (e.g., 16 bytes for AES-CTR), but we are forced to load and store whole blocks (typically 4096 bytes).

We avoid this problem by caching IV blocks in RAM, in an LRU cache of predefined depth (e.g., 1024 entries). For the performance reasons just discussed, this cache is not write-through. This way, we coalesce possibly many updates of the same IV block (triggered by many logical {\SflcWrite}s to the same data block, or by logical {\SflcWrite}s to many blocks within the same slice) into just one physical \bWrite, thereby lowering the I/O overhead.

For each physical slice, we pack the IVs for the blocks contained therein into $\deltas$ physical blocks at the beginning of the slice. There is a simple static correspondence between a physical block and the on-disk location of its IV: the $(j + \deltas)$-th block within a physical slice is encrypted by the $j$-th IV within the initial $\deltas$ blocks. Hence, we assume the existence of functions $\LoadIV$ and $\SampleAndStoreIV$ (with self-explanatory behaviour) which take as input a physical address $\Bphy$ and return the corresponding IV. Analogously, we consider the functions $\Encrypt(\ptxt,\IV,\key)$ and $\Decrypt(\ctxt,\IV,\key)$ as acting on blocks according to the implied mode of operation.

\paragraph{Indirection layer.} Consider a read or write operation to a logical block address $\Blog$ for volume $\V_i$. There are three possible cases:

\begin{enumerate}
\item The requested operation (read or write) happens on a block whose logical address $\Blog$ was previously allocated. In this case we need to map efficiently $\Blog$ for volume $\V_i$ to the corresponding physical address $\Bphy$ on the device.
\item It is a $\SflcRead$ operation for a $\Blog$ which falls within a slice that was not allocated before. We need to specify the behavior in this case.
\item It is a $\SflcWrite$ operation for a $\Blog$ which falls within a slice that was not allocated before. We need to define how to allocate a new physical slice.
\end{enumerate}

Notice that the offset of a block \emph{within} a slice is left unchanged by the slice map. So, we can find the LSI $\LSI$ to which $\Blog$ belongs as simply $\LSI := \lfloor \Blog / \Sl \rfloor$. Then we need to check whether $\SliceMap_i[\LSI]$ is defined or not. Luckily, as we have seen above, slice maps are not too large. So, when decrypting the slice maps during device instantiation, we can store in memory a full view of these slice maps for each volume as arrays of PSIs indexed by the related LSI, and define a special return symbol $\bot$ for those LSIs which have not yet been assigned.

Then let's analyse the three cases above one by one. In the first case, we just need to consider the PSI $\PSI$ just obtained and add a correct offset. This will give us the physical address $\Bphy$ where we will find the data to decrypt. From what we just discussed, this can be done as $\Bphy := \PSI \cdot \Sp + \deltas + (\Blog \mod \Sl)$.

In the second case, instead, $\SflcRead$ should return a default, non-error value, e.g., $0$. Not throwing an error in this case is necessary for the semantic of a volume: Although it has never been written before, the block logically exists, it is within the logical boundary of the volume, so it would be incorrect to return an error. Notice how we never allocate new slices on $\SflcRead$ requests: as we will see in~\expref{Section}{sec:security}, this is necessary for security, and prevents logical read operations to leave a trace on the disk.

In the third and last case we need a way to allocate a new slice for volume $\V_i$ at a position consistent with $\Blog$. We do this through a function $\NewSlice$ which returns a PSI uniformly at random among those not yet mapped. We will see in the next paragraph how to implement this function.

\begin{figure}
    \centering
    \begin{minipage}{.48\linewidth}
\begin{algorithm}[H]
\captionsetup{font={footnotesize}}
\caption{\SflcRead($\V_i, \Blog$)}\label{alg:sflcread}
\begin{algorithmic}[1]\footnotesize
\State $\LSI := \lfloor \Blog / \Sl \rfloor$
\State $\PSI \leftarrow \SliceMap_i [ \LSI ]$
\State \textbf{if} $\PSI = \bot$
\State ~~~~ \Return $0$
\State $\Bphy := \PSI \cdot \Sp + \deltas + (\Blog \mod \Sl)$
\State $\IV \leftarrow \LoadIV(\Bphy)$
\State $\ctxt \leftarrow \bRead(\Bphy)$
\State \Return $\Decrypt(\ctxt, \IV, \VEK_i)$
\end{algorithmic}
\end{algorithm}
    \end{minipage}
    \hfill
    \begin{minipage}{.48\linewidth}
\begin{algorithm}[H]
\captionsetup{font={footnotesize}}
\caption{\SflcWrite($V_i, B, d$)}\label{alg:sflcwrite}
\begin{algorithmic}[1]\footnotesize
\State $\LSI := \lfloor \Blog / \Sl \rfloor$
\State $\PSI \leftarrow \SliceMap_i [ \LSI ]$
\State \textbf{if} $\PSI = \bot$
\State ~~~~ $\PSI \leftarrow \NewSlice(\V_i, \LSI)$
\State ~~~~ \textbf{if} $\PSI = \bot$ \Return \texttt{Error}
\State $\Bphy := \PSI \cdot \Sp + \deltas + (\Blog \mod \Sl)$
\State $\IV \leftarrow \SampleAndStoreIV(\Bphy)$
\State $\ctxt := \Encrypt(d, \IV, \VEK_i)$
\State \Return $\bWrite(\Bphy, \ctxt)$
\end{algorithmic}
\end{algorithm}
    \end{minipage}
\end{figure}

\paragraph{Slice allocation.} slice mappings are created \emph{lazily}, only when the first request for a block belonging to a new, yet-unmapped slice arrives. At this point, we create the mapping for this slice by sampling a physical slice uniformly at random \emph{among the free ones}: this guarantees that no conflicts arise between volumes, and their slices end up randomly interleaved on the disk. To make this lazy sampling possible, we need to implement efficiently a function \NewSlice: given as input a volume's identifier $\V_i$ and an LSI $\LSI$, it returns a corresponding PSI $\PSI$ for the new slice. There are different ways to implement this, here we give a reference description using a permutation of the array representing the slices.

Concretely, we keep an in-memory, per-device array $\prmslices$ of PSIs of size at most $\maxslices = \lceil \frac{\devbsize-\Shdr}{\Sp} \rceil$ (maximum possible number of slices), and an array of the same size \ofld, an \emph{``occupation bitfield''} telling us which physical slices are occupied (the PSIs are the indexes of the array). Initially, at device instantiation, $\prmslices$ is initialized as $\prmslices[i] := i$ and then permuted using the efficient Fisher-Yates algorithm~\cite{fisheryates}. The bitfield \ofld is initialized with all \free values. 
When a volume is {\sflcopen}ed, and we discover the physical slices it maps to, we mark them as \occupied in \ofld. The slice allocation algorithm then simply works by repeatedly taking the next element from the pre-shuffled array of PSIs until the bitfield tells us it is free (at which point we take it and mark it as occupied). A stateful occupation counter $\octr$ is kept to facilitate this task, initially set to $0$, and then increased up to the first element of $\prmslices$ marked as $\free$ (i.e., the first $\octr$ elements of $\prmslices$ are guaranteed to be $\occupied$ in \ofld). This way, when a new slice is required, it is immediate to advance to the next available one. We also need to update $\V_i$'s slice map $\SliceMap_i$ before returning (this is so far the only change that will eventually persist, encrypted, on disk).

\begin{algorithm}[H]
\captionsetup{font={footnotesize}}
\caption{\NewSlice($\V_i, \LSI$)}\label{alg:NewSlice}
\begin{algorithmic}[1]\footnotesize
\State \textbf{while} $\ofld[\prmslices[\octr]] = \occupied$ \textbf{do:} $\octr:=\octr+1$
\State $\PSI_\mathsf{new} \from \prmslices[\octr]$
\State $\ofld[\prmslices[\octr]] := \occupied$
\State $\SliceMap_i[\LSI] := \PSI_\mathsf{new}$
\State \Return $\PSI_\mathsf{new}$
\end{algorithmic}
\end{algorithm}

\vspace{-0.5cm}

The permutation of indexes changes at every device instantiation, so that the mapping is not static. The size of these in-memory supporting data structures (array and bitfield) is $O ( \frac{\devbsize}{\Sp} \log \frac{\devbsize}{\Sp} )$, as usual. The lazy allocation technique is also what allows us to \emph{overcommit} the total physical storage space: we can have the sum of the sizes of the logical volumes exceed the total physical storage space, as long as the sum of ``actually used'' spaces does not. 
However, it might be useful to intentionally limit this overcommitment (for the opened volumes) to decrease the risk of I/O errors and improve user experience. This can be done using the metadata field in the VMB ciphertexts, as explained in~\expref{Section}{sec:metadata}.

\paragraph{Handling slice conflicts.} Unsafe user operations could lead to accidental \emph{corruption} of one or more volumes, resulting in the same PSI being assigned to different LSIs on different slice maps (see, e.g.,~\expref{Section}{sec:opmod}). This would create a persistent ambiguity on volume instantiation, which would hamper any attempts from the upper layer (for example, as explained in~\expref{Section}{sec:corruptionresistance}) to restore data content. To avoid this, Shufflecake resolves any slice allocation conflict by duplicating and reassigning corrupted slices for every affected volumes. To be more precise:

\begin{itemize}

\item When reading an entry $\PSI$ from $\SliceMap_i[\LSI]$ for an LSI $\LSI$, a collision for $\PSI$ is detected by seeing that the current element being initialised $\ofld[\PSI]$ is already set to \occupied. That means that the physical slice with index $\PSI$ has been corrupted: formerly, it belonged to volume $\V_i$ (and, possibly, that was also corrupted from a volume of even higher order), but it has been mistakenly assigned also to a volume $\V_j$ for some unknown $j < i$.

\item When this happens, do not attempt to resolve the collision by modifying $\SliceMap_j$. First, because by the time of this detection we might not know $j$, and second, because it might break plausible deniability in the multi-snapshot setting: If the corruption happens because an adversary has forced the user to open and write data to $\V_j$ (as explained in~\expref{Section}{sec:opmod}, then the user is left alone and tries to repair the slice collision by changing $\SliceMap_j$, later at the next inspection the adversary sees that the content of $\V_j$ has not changed, and yet $\SliceMap_j$ has, which would not be plausibly explainable.

\item Instead, the solution is to modify $\SliceMap_i$ by allocating a new slice with $\NewSlice(\V_i , \LSI)$, which also sets $\SliceMap_i[\LSI]$.

\end{itemize}

This has the disadvantage that the corruption of a few blocks due to operations performed on $\V_j$ makes the whole slice unreadable by $\V_i$ (because the freshly assigned slice contains garbage). However, remediating volume corruption is not a primary goal of Shufflecake, as this can only happen as a consequence of unsafe volume operations (see~\expref{Section}{sec:opmod}). The important point here is that this mechanism avoids the problem of allocation ambiguity persisting on the slice maps, thereby giving some hope for upper layer tools to repair the corrupted volumes. Possible improvements on this mechanism are described in~\expref{Section}{sec:corruptionresistance}. There, in particular, it is argued that by carefully \emph{cloning} the content of the corrupted slice to the newly assigned slice, one can hope to recover from corruption at the block granularity rather than the slice granularity.

\paragraph{Device instantiation.} We are now ready to explain in detail how the process of device instantiation works. 
We assume a subroutine \HandleCorruption: this is the procedure responsible for handling the case of ambiguous slice assignment derived from a volume corruption. As explained above, in this reference description the corruption is handled by simply reassigning a new slice to the topmost corrupted volume. 
The full instantiation procedure is depicted in~\expref{Algorithm}{alg:instantiation}. Notice that volume information is recovered `backward', i.e., VMBs are decrypted starting from $\Nvol-1$ down to $0$. However, once all VMBs (and related slice maps) are decrypted, the volumes are {\sflcopen}ed in forward order, meaning that $\ofld$ is populated by reading content from $\SliceMap_0$ to $\SliceMap_{\Nvol-1}$.

\begin{algorithm}[H]
\captionsetup{font={footnotesize}}
\caption{\Instantiate(\dvc, \pwd)}\label{alg:instantiation}
\begin{algorithmic}[1]\footnotesize
\State \textbf{read} \salt from \dvc
\State $\KEK := \KDF(\pwd,\salt)$
\State \textbf{for} $j=0,\ldots,\maxvols-1$
\State ~~~~ \textbf{read} $(\IV,\ctxt)$ from $\dvc.\DMB.\cell[j]$
\State ~~~~ $\mathsf{topvmk} := \Decrypt(\ctxt, \IV , \KEK)$
\State ~~~~ \textbf{if} $\mathsf{topvmk} \neq \bot$ \textbf{then exit for} \Comment{Password correct: top volume found}
\State \textbf{next} $j$
\State \textbf{if} $j=\maxvols$ \Return ``ERROR: wrong password.''
\State $\Nvol := j+1$
\State global arrays \VMK, \VEK, \metadata of size $\Nvol$, global var \numslices
\State $\VMK_{\Nvol-1} := \mathtt{topvmk}$
\State \textbf{for} $i=\Nvol-1,\ldots,0$
\State ~~~~ \textbf{read} $(\IV,\ctxt)$ from $\dvc.\VMB_i$
\State ~~~~ $(\VEK_i, \mathsf{prevVMK}, \mathsf{thisnumslices}, \metadata_i) := \Decrypt(\ctxt, \IV, \VMK_i)$
\State ~~~~ \textbf{if} $i \neq 0$ \textbf{then} $\VMK_{i-1} := \mathsf{prevVMK}$
\State ~~~~ \textbf{if} $i = \Nvol-1$ \textbf{then} \Comment{Sanity checks}
\State ~~~~ ~~~~ \textbf{if} $\mathsf{thisnumslices} > \lceil \frac{\devbsize-\Shdr}{\Sp} \rceil$ \textbf{then} \Return ``ERROR: oversize volume.''
\State ~~~~ ~~~~ $\numslices := \mathsf{thisnumslices}$
\State ~~~~ \textbf{end if}
\State ~~~~ \textbf{if} $\numslices \neq \mathsf{thisnumslices}$ \textbf{then} \Return ``ERROR: size mismatch.''
\State ~~~~ global array $\SliceMap_i$ of size $\numslices$
\State ~~~~ \textbf{read} $\IV'$ and $\ctxt'$ of size $\numslices$ from blocks after $\dvc.\VMB_i$
\State ~~~~ $\SliceMap_i := \Decrypt(\ctxt', \IV', \VEK_i)$
\State \textbf{next} $i$
\State global arrays \prmslices, \ofld of size \numslices, global var \octr
\State $\octr := 0$
\State \textbf{for} $i = 0,\ldots, \numslices-1$
\State ~~~~ $\prmslices[i] := i$
\State ~~~~ $\ofld[i] := \free$
\State \textbf{next} $i$
\State \textbf{for} $i = 1,\ldots, \numslices-1$ \Comment{Fisher-Yates shuffle}
\State ~~~~ $j \fromunif \set{0,\ldots,i}$
\State ~~~~ $\mathtt{swap}(\prmslices[i], \prmslices[j])$
\State \textbf{next} $i$
\State \textbf{for} $i = 0,\ldots, \Nvol-1$ \Comment{Scan across all volumes}
\State ~~~~ \textbf{for} $\LSI = 0, \ldots, \numslices-1$ \Comment{Scan all slice map of $\V_i$}
\State ~~~~ ~~~~ $\PSI \from \SliceMap_i[\LSI]$
\State ~~~~ ~~~~ \textbf{if} $\ofld[\PSI] = \occupied$ \textbf{then}
\State ~~~~ ~~~~ ~~~~ $\HandleCorruption(\V_i, \PSI, \LSI)$ \Comment{Corruption found; repair}
\State ~~~~ ~~~~ \textbf{else}
\State ~~~~ ~~~~ ~~~~ $\ofld[\PSI] := \occupied$
\State ~~~~ ~~~~ \textbf{end if}
\State ~~~~~ \textbf{next} $\LSI$
\State ~~~~~ \textbf{create} virtual block device $\V_i$ of size $\Sl \cdot \numslices$
\State \textbf{next} $i$
\State \Return
\end{algorithmic}
\end{algorithm}

\subsubsection{Volume actions.} In principle, nothing in the scheme inherently prevents us from creating, opening, and closing volumes freely and independently, at any time. However, for real-world operations, we force volumes to be opened in a hierarchical way, by only providing \emph{one} password (for the most secret volume).

To create a new volume it is needed: the index $i$ of the new volume $\V_i$, the chosen password, and the VMB key of volume $\V_{i-1}$ (if $i > 1$). This way, one can format the header by generating the relevant keys, filling the $\VMK_{i-1}$ field, and initialising the slice map as empty. No operation is needed on the data section.

To open a volume, only its password is needed in order to decrypt the header, which then allows to load its slice map and to decrypt its slices. Finding the right header for a provided password is done simply by trying every one of them, until the authenticated ciphertext in the related DMB cell decrypts correctly.

Closing a volume mainly modifies the state of the Shufflecake instance in RAM, by removing the relevant volume information (and securely erasing its key). The only required disk operations are the ones needed to persist some possibly-unsynchronised data.

No specific operation is needed to destroy a volume, i.e. to remove it from the disk. It is enough to just forget the password, or to overwrite the header with random bytes: by the PD guarantees, there is no way to then even prove that there was a volume in that slot, let alone to decrypt it.

\subsection{Operational Model}\label{sec:opmod}

In this section, we define the operational model of Shufflecake, to provide a safe \emph{mode of use} allowing the user to retain both plausible deniability and data integrity. Besides some general constraints, we specify what the user has to do in ordinary working conditions, and how instead they must behave when confronted with the adversary.

\subsubsection{Risk of data corruption.} A simple observation shows how a legitimate-looking usage mode of Shufflecake actually entails a high risk of data corruption. If we do not open all $\Nvol$ existing volumes, and instead only open the ones we plan to use, we do not load all the slice maps in RAM, which leads to an incorrect reconstruction of the complete device's bitfield \texttt{ofld} of free physical slices. The physical slices belonging to the still-closed volumes will be counted as free, and will therefore possibly be allocated to the open volumes during data write, which would then overwrite their content. This can only be avoided with certainty by always opening \emph{every} volume, regardless of which ones we are going to use: if the password to a volume is not provided, Shufflecake has no way of detecting its existence. It then follows from the overcommitment of the physical storage space that we risk re-using its physical slices for some other volume.

However, mitigation does not need to be perfect. It could be possible to reduce the risk of corruption when not opening all volumes by using some form of \emph{error-correction} on the unopened volumes (hence sacrificing some space), and then trying to recover the volume if corruption happens. We discuss this in~\expref{Section}{sec:corruptionresistance}.

\subsubsection{General constraints.} The first thing to do when initialising a device with Shufflecake is to fill it completely with random bytes. Though long and tedious, this operation is crucial even for single-snapshot security, as we will see in~\expref{Section}{sec:security}, just like it is for TrueCrypt. The most sensitive data should be placed in a volume of sufficiently high order. We cannot, of course, give precise indications of the form ``use at least 3 volumes'', or ``6 volumes should be safe enough'', because, by Kerchoff's principle, we assume that the adversary knows about Shufflecake, and in particular reads this document. The volumes of lower order, that will be disclosed to the adversary, should be filled with ``mildly incriminating'' data, so as to convince the attacker that one had a plausible reason to hide them. We do not specify more precisely what kind of content would be suited to this end, partly for the same reasons as before (the adversary would immediately flag it as decoy content, and ask for more passwords), partly because it heavily depends on the context and on who the adversary concretely is. The decoy volumes must also be otherwise ``credible'', in particular they must be formatted with realistic file systems, and they must be reasonably up to date. Periodic updates can be delegated to a background daemon or offloaded to the user.

\subsubsection{Home alone.} In normal operating conditions, when not confronted with the adversary, the recommended course of action for the user is to unlock all the volumes present on the device, in order to prevent data corruption as explained before. The reason behind the design choice of chaining the volumes into a linked list to help the user in that regard: this way, the user is able to open all volumes by just providing the password to $\V_{\Nvol-1}$. In our implementation, this is actually the mandated semantic of the \sflcopen operation: the user only provides the password of the \emph{last} volume they want to open, the previous ones are {\sflcopen}ed automatically. Other implementations might opt instead to ignore the $\VMK_{i-1}$ field in the volume header, and give more flexibility to the user, if aware of the risks entailed.

\subsubsection{Under interrogation.}\label{sec:interrogation} When questioned by the adversary and forced to reveal passwords, the user must obviously not surrender more than $\Nvol - 1$ of them (otherwise there would be nothing left to protect). Although irrelevant for the cryptographic security of the scheme, we stress that, in order for the user's lie to be credible, they must only reveal the 
decoy passwords under a certain amount of pressure, or after some time has passed. Notice that a responsible and safe use of Shufflecake puts on the user the burden of being able recall quickly and reliably these decoy passwords, even under distress. This might be hard to get right given that the recommended course of action for the user in daily use is to only open the most hidden volume. It is up to the user to define the maximum number of volumes $\Nvol$ that makes them comfortable in this task. Shufflecake implementations might include additional features to aid the user in this sense, for example a function to check the password of a decoy volume without actually opening it, or even a puzzle which, with some random probability, prompts the user to also insert the password for a decoy volume when opening a hidden one.

\subsubsection{Safeword.}\label{sec:safeword} As we previously discussed, one big operational difference between Shufflecake (but also other solutions like, e.g., HiVE) and TrueCrypt is that with Shufflecake ``the adversary does not know when she can stop questioning you'', because there is no way to prove that a given password unlocks all the existing volumes on a given device (unless it's the password unlocking the $(\Nvol-1)$-th volume). In TrueCrypt, instead, there is either just a regular volume, or a decoy and a hidden volume. This distinction might be important in those scenarios where a user wants to avoid the possibility of looking uncooperative to a certain adversary. In such scenarios, the user might want to have the choice of surrendering all the volumes, and a method to convince the adversary that no other volumes exist, possibly using an additional ``full disclosure password'' that we will call \emph{safeword}.

One simple way to implement this, even in Shufflecake, is to actually create all \maxvols volumes when initializing a device. Clearly, remembering \maxvols passwords would be quite cumbersome for the user, so the solution is to only remember $\Nvol+1$ passwords instead: those for the $\Nvol$ volumes that are actually desired, and the extra one for the last {\maxvols}-th volume, which is going to be the safeword. In fact, the user will never need to open more than $\Nvol$ volumes for regular use. As discussed in~\expref{Section}{sec:opmod}, this might harm the consistency of the other, unopened volumes, but this is not important: by using the safeword, the user would still be able to convince the adversary that all volumes have been revealed due to the linkage between them and the ciphertext authentication in the headers. Analogously, in solutions like TrueCrypt, a simple way to implement a safeword is to actually always create a hidden volume, even if only a standard volume is desired.

We stress, however, that using this feature is a dangerous proposition: If such possibility exists, and users are allowed to do that, then why not to? The adversary might arguably assume that a user \emph{must} have a safeword, and pressure for its disclosure. This would put at risk those users who decide to not use this feature, who might then be pushed to its adoption. This, in turn, would ruin plausible deniability for everyone, because now we have a system where everyone has a safeword by default.

We believe there is no simple solution to this dilemma: One has either to accept the risk of looking uncooperative and be subject to further interrogation, or to give up PD at all. We remark that, as far as we know, the issue of a safeword feature (or even just its possibility) for plausible deniable filesystems has not been addressed in the literature before, as all implementations we are aware of (including woORAM-based ones) employ some form of architectural hard limit on the number of possible nested levels of secrecy. We believe this to be a serious operational problem for the security of PD solutions. For this reason, not only do we discourage the use of this feature, but we also propose a way to make the implementation of \emph{any} safeword-like system impossible. This is discussed in~\expref{Section}{sec:unbound}, and boils down to the idea of having an unbounded number of possible volumes per device.

\subsection{Security}\label{sec:security}

In this section, we prove that (`Legacy') Shufflecake achieves single-snapshot security, as defined in \expref{Section}{sec:ss}. 

\begin{theorem}[Single-snapshot security of Shufflecake `Legacy']\label{thm:sflcsec}
The Shufflecake `Legacy' scheme as described in \expref{Section}{sec:sflcdesign} is a single-snapshot (SS) secure PD scheme according to \expref{Definition}{def:ss}.
\end{theorem}

\subsubsection{Assumptions.}\label{sec:secassumptions} 
In proving \expref{Theorem}{thm:sflcsec} we will make some assumptions in order to keep the proof compact and intuitive. We assume w.l.o.g. that all passwords are encoded as bitstrings of length \secpar. Notice how throughout all \expref{Section}{sec:sflcdesign} we avoided giving concrete security parameters. Although in the real-world instantiation of Shufflecake we are going to have cryptographic primitives with input and output of fixed size (e.g., 128-bit IVs, 256-bit keys, etc), in the context of this proof we can consider them of variable length. This will allow us to produce an asymptotic bound, and to apply it as in \expref{Definition}{def:ss} in order to prove that the advantage of any (computationally bounded) adversary is indeed negligible in the security parameter. In so doing, we will treat the cryptographic primitives used in the Shufflecake design as \emph{ideal}. More specifically:

\begin{itemize}
    \item The KDF will be replaced by a \emph{random oracle} $\oracle_K$, mapping $\secpar$-bit passwords and salts to truly-random $\secpar$-bit strings.
    \item The symmetric encryption scheme will be replaced by an \emph{ideal cipher} $\escheme$, mapping $\secpar$-bit keys and IVs to truly-random permutations over $\set{0,1}^\secpar$.
    \item The authenticated encryption used in the DMB will be replaced by a pair of oracles: the oracle $\oracle_{AE}$, mapping $\secpar$-bit keys and IVs to truly-random injections between plaintext and ciphertext spaces, and its inverse which returns a constant $\bot$ failure symbol if queried outside of the codomain.
\end{itemize}

All these oracles will be initialised by the game and provided to the adversary \adver to be queried freely (at a unitary time cost).

\subsubsection{The proof.}\label{sec:proof}
Let us consider \expref{Experiment}{exp:pd} under the constraints explained in \expref{Section}{sec:ss}, and let $D$ be the ``challenge disk snapshot'' provided to the adversary \adver by the game (i.e., either $D_0$ or $D_1$ according to the secret bit $b$). For the given \adver, we will consider $\Nvol$, the decoy passwords $P_0, ... , P_{\Nvol-2}$, and the access patterns $O_0$ and $O_1$ as public parameters of the game instance.

Let us first notice that all the oracle queries performed by \adver \emph{before} receiving the challenge disk snapshot $D$ cannot change \adver's advantage, because they are completely uncorrelated with $D$ (and the secret bit $b$). We can, therefore, safely disregard those queries in our analysis.

Then, let us define $Q$ to be the ordered sequence of all queries $\set{q_i}_i$ made by \adver to the oracles \emph{after} receiving the challenge $D$, and define $n := |Q|$. Also define $Q_i$ to be the sequence of queries $(q_0, \ldots, q_{i-1})$ up to the $i$-th query. Analogously, let us define $R$ to be the ordered sequence of all responses $\set{r_i}_i$ returned by the oracles; also define $R_i$ to be the sequence of responses $(r_0, \ldots, r_{i-1})$ up to the $i$-th response. We consider the execution of \adver as the execution of a sequence of single-query stateful adversaries, where the state is just the `history' of the previous queries: $\adver_0\left(D\right) \to q_0$, $\adver_1\left(D, Q_0, R_0\right) \to q_1$, and so on until $\adver_{n-2}\left(D, Q_{n-2}, R_{n-2}\right) \to q_{n-1}$ and $\adver_{n-1}\left(D, Q_{n-1}, R_{n-1}\right) \to b'$.

Let $\KEK_i$, $\VMK_i$ and $\VEK_i$ be, respectively, the key-encryption key, the VMB key, and the volume encryption key of volume $\V_i$. Rigorously speaking, in the security game, the values $P_{\Nvol-1}, \KEK_{\Nvol-1} , \VMK_{\Nvol-1}$, and $\VEK_{\Nvol-1}$ are only sampled if $b = 0$. Let us instead consider them to be sampled anyway, and left unused in the case $b = 1$. We denote by $\mathcal{S}$ the tuple $(P_{\Nvol-1}, \KEK_{\Nvol-1}, \VMK_{\Nvol-1}, \VEK_{\Nvol-1}) \in \set{0,1}^{4\secpar}$. Let us define the event $E_i$ as the event that either of $P_{\Nvol-1}$, $\KEK_{\Nvol-1}$, $\VMK_{\Nvol-1}$, or $\VEK_{\Nvol-1}$ appear in query $q_i$ (we say that query $q_i$ \emph{strikes}). Finally, let us define $E := E_0 \cup \ldots \cup E_{n-1}$ the event that \emph{at least one} query strikes. We will first prove two lemmata.

\begin{lemma}\label{lem:1}
$\Pr{E} = \Pr{E \,|\, b=0} = \Pr{E \,|\, b=1} = \negl(\secpar)$. 
\emph{(The adversary can only guess one of the secrets of $\V_{\Nvol-1}$ with negligible probability.)}
\end{lemma}
\begin{proof}
Let us prove that $\mathcal{S}$ is \emph{statistically independent} from any query $q_i \in Q$.

This tuple gets sampled uniformly at random from $\set{0,1}^{4\secpar}$, so its distribution does not depend on the public values. Since the oracles implement \emph{ideal} cryptographic primitives, it follows that their outputs are statistically independent from their inputs; this does not just hold \emph{marginally} for single input-output pairs, but \emph{jointly}: any tuple of inputs is statistically independent from the tuple of corresponding outputs (for the ideal cipher $\escheme$, this only holds for the \emph{key} inputs). In particular, $\mathcal{S}$ is statistically independent from the whole tuple $(D, R)$. Since $\adver_i$'s inputs, namely the tuple $(D, Q_{i-1}, R_{i-1})$, are a (randomised) function of $(D, R)$, we deduce that its output, namely $q_i$, must also be independent of $\mathcal{S}$.

Therefore, since $(P_{\Nvol-1}, \KEK_{\Nvol-1}, \VMK_{\Nvol-1}, \VEK_{\Nvol-1})$ are uniform, the probability that $q_i$ contains, say, $P_{\Nvol-1}$, is $2^{-\secpar}$.
By the union bound, $\Pr{E_i} \leq 4 \cdot 2^{-\secpar}$. Using the union bound again, we get:

\begin{align*}
\Pr{E} \leq \sum_{i=0}^{n-1} \Pr{E_i} \leq n \cdot 4 \cdot 2^{-\secpar}
\end{align*}

This expression is clearly $\negl(\secpar)$, since $n$ must be at most a polynomial in $\secpar$. The final claim on the equality of the conditional probabilities follows by simply observing that this reasoning holds irrespective of the value of $b$.\qed
\end{proof}

\begin{lemma}\label{lem:2}
$\Pr{\adver\left(D\right) \to 1 | b=0 \wedge \bar{E}} = \Pr{\adver\left(D\right) \to 1 | b=1  \wedge \bar{E}}$. 
\emph{(Unless she can guess one of $\V_{\Nvol-1}$'s secrets, the adversary has the exact same \emph{view} in the cases $b=0$ and $b=1$.)}
\end{lemma}
\begin{proof}
It is sufficient to prove that $\adver_{n-1}$'s inputs, namely $D$, $Q_{n-1}$, and $R_{n-1}$, follow the same \emph{joint conditional distribution}, conditioned to the event $\bar{E}$, regardless of whether $b = 0$ or $b = 1$. 

We prove this by induction. Using a concise notation, we want to prove that the following quantity does not depend on the bit $b$:

\begin{align*}
\Pr{D, Q_{n\mbox{-}1}, R_{n\mbox{-}1} | \bar{E}, b} \!=\! \Pr{q_{n\mbox{-}1}, r_{n\mbox{-}1} | D, Q_{n\mbox{-}2}, R_{n\mbox{-}2}, \bar{E}, b} \!\cdot\! \Pr{D, Q_{n\mbox{-}2}, R_{n\mbox{-}2} | \bar{E}, b}
\end{align*}

Showing that the first factor is independent of $b$ will prove the inductive step. To this end, let us further rewrite it as:

\begin{align*}
\Pr{q_{n-1} | D, Q_{n-2}, R_{n-2}, \bar{E}, b} \cdot \Pr{r_{n-1} | q_{n-1}, D, Q_{n-2}, R_{n-2}, \bar{E}, b}
\end{align*}

The first factor is independent of $b$ because $q_{n-1}$ is the output of $\adver_{n-2}$, which only takes $D, Q_{n-2}, R_{n-2}$ as inputs, all of which are among the conditioning terms already. The second factor is independent of $b$ because, given that $\bar{E}$ holds, the oracles behave the same whether $b = 0$ or $b = 1$. This is because striking queries are the only oracle inputs that could trigger responses with unequal distributions (i.e., correlated to $D$) in the cases $b=0$ and $b=1$. But if, instead, we rule these queries out by conditioning on $\bar{E}$, then the oracles instantiated when $b=0$ are perfectly interchangeable with the ones instantiated when $b=1$.

We are now left to prove the base step for induction, corresponding to $ \Pr{D \,|\, \bar{E}, b} $. Let us rewrite it, using Bayes' theorem, as:

\begin{align*}
\Pr{D | \bar{E}, b} = \frac{\Pr{\bar{E} | D, b} \cdot \Pr{D | b}}{\Pr{\bar{E} | b}}
\end{align*}

The term $\Pr{\bar{E} | b}$ is independent of $b$ by \expref{Lemma}{lem:1}. The same is true of $\Pr{\bar{E} | D, b}$: the same proof applies as for \expref{Lemma}{lem:1}, because the reasoning is unchanged when we condition the probabilities on a particular realisation for $D$.

We only have to prove that $\Pr{D | b}$ does not depend on the bit $b$, i.e. that the disk snapshot follows the same \emph{a-priori} (non-conditioned) distribution, whether $b = 0$ or $b = 1$. To prove this, we will use the actual properties of the Shufflecake scheme. Let us proceed by analysing the disk layout region by region.

The blank spaces (empty DMB cells, unmapped slices, etc.) are filled with equally-distributed uniformly-random noise. The spaces occupied by volume $V_{\Nvol-1}$, when $b=0$, are filled with oracle responses to queries containing one of $V_{\Nvol-1}$'s secrets. These responses are fresh randomnesses, which follow the same distribution as the noise filling the same spaces when $b=1$. 

We are only left to prove that the (decrypted) logical contents and metadata of the decoy volumes follow the same distribution when $b=0$ and when $b=1$. Indeed, the \emph{logical} contents of the volumes are fixed and identical in the two cases, determined by $O_0$ and $O_1$ (this is because they have to follow the constraint defined in \expref{Section}{sec:prelimsecpd}). Also, the DMB cells and the VMBs contain equally-distributed uniformly-random oracle outputs (keys, etc.).

The last step is to show that the slice maps of the decoy volumes follow the same distribution in the two cases $b=0$ and $b=1$. By the second constraint on the access pattern, defined in~\expref{Section}{sec:prelimsecpd}, we get that slice allocation is triggered for the same LSIs of the decoy volumes in both the two cases. Even though some more slice allocations are performed on $V_{\Nvol-1}$'s LSIs when $b=0$, this does not impact the resulting observable distribution on the PSIs assigned to decoy LSIs. This is because slice allocation always takes a PSI randomly among the free ones, therefore the order in which the LSIs are mapped can be permuted freely without impacting the distribution. Thus, even in the case $b=0$, we can equivalently imagine that the decoy LSIs get all mapped before $V_{\Nvol-1}$'s ones, yielding the same distribution as when $b=1$. This concludes the proof.\qed
\end{proof}

\begin{proof}[Proof of \expref{Theorem}{thm:sflcsec}]
We use \expref{Lemmata}{lem:1} and \ref{lem:2} to prove that the advantage of \adver, as defined in \expref{Equation}{eqn:1} of \expref{Definition}{def:ss}, is negligible. By conditioning both terms of \expref{Equation}{eqn:1} to the events $E$ and $\bar{E}$, we get: 

\begin{align*}
&\Big| \Pr{\adver\left(\mathsf{D}\right) \!\to\! 1 | b\!=\!0} - \Pr{\adver\left(D\right)\!\to\!1 | b\!=\!1} \Big| = \\
=& \Big|
   \Pr{E} \Big( \Pr{\adver\left(D\right) \!\to\! 1 | b\!=\!0 \wedge E} -
    \Pr{\adver\left(D\right) \!\to\! 1 | b\!=\!1 \wedge E} \Big) + \\
+& \Pr{\bar{E}} \Big( \Pr{\adver\left(D\right) \!\to\! 1 | b\!=\!0 \wedge \bar{E}} - \Pr{\adver\left(D\right) \!\to\! 1 | b\!=\!1 \wedge \bar{E}} \Big) \Big| \leq \\
\leq& \Pr{E} \cdot 1 + \Pr{\bar{E}} \cdot 0 = \mathsf{negl}(\secpar),
\end{align*}
which concludes the proof.\qed
\end{proof}

\section{Implementation and Benchmarks}\label{sec:implementation}

We implemented the (`Legacy') Shufflecake scheme in the C language as an open-source device-mapper-based driver for the Linux kernel. We published our code under the GPLv2+ license. The current release is \sflclastversion~\cite{sflcwebsite}. This section describes the programming environment and the structure of our implementation, and presents concrete performance measurements taking other popular disk encryption solutions as a baseline for comparison.

\subsection{Structure of the Implementation}\label{sec:implstructure}

Our implementation consists of two components: a \dmsflc kernel module (which does most of the job), and a \shufflecake companion userland application (used to correctly manage the volumes). The kernel module is the component that actually implements the scheme, translating logical requests into physical requests, and persisting slice maps into the respective headers.

\subsubsection{Cryptography.} Cryptographic primitives are provided by the Libgcrypt library~\cite{libgcrypt}. We target 128 bits of security. We use Argon2id as a KDF, which was implemented in Libgcrypt recently~\cite{libgcryptargon}. We use AES-GCM-256 as an authenticated cipher, and AES-CTR-256 for data encryption, with 128-bit IVs.

\subsubsection{The userland application.}\label{sec:userland} This component is used to manage volumes creation, opening, and closing. To this end, it manages the DMB and VMB of each volume header. The $\VEK_i$ is passed to the kernel module to decrypt the slice map and data section of that volume, while $\VMK_{i-1}$ is used to iteratively open all the less-secret volumes, as described in \expref{Section}{sec:sflcdesign}.

This is offloaded to the userland application because key management is arguably better handled in user space: for example, we need to react to an incorrect password by asking the user to try again, not by emitting a kernel log message. There is also another technical hindrance to delegating everything to the kernel module: state-of-the-art KDFs like Argon2id~\cite{argon2} are not currently implemented in the Linux Kernel Crypto API~\cite{kernel-crypto}, while they are available in user-space software libraries like Libgcrypt~\cite{libgcrypt}.

The other blocks of the volume header, which contain the slice map encrypted with the $\VEK$, are managed by the kernel module (except at \sflcinit time, when an empty slice map is written by the userland tool).

\subsubsection{Volume operations.} The \shufflecake \sflcinit command takes as input a device path, and then interactively asks the user a number $\Nvol \leq \maxvols$ and $\Nvol$ passwords as input, correctly formats the first $\Nvol$ volume headers, and fills the remaining $\maxvols - \Nvol$ slots with random bytes; this way, the pre-existing volumes are wiped by erasing their headers (crypto-shredding). Unless a \texttt{--skip-randfill} option is provided (e.g., for testing or debugging purpose), the whole disk is filled with random bytes before formatting the header section. This command only formats the disk: it does not create the Linux virtual devices associated to the volumes.

The \shufflecake \sflcopen command takes a device path as input and asks one single password to the user, then looks up the volume headers, and opens all the volumes starting from the one whose password is provided, backwards up to the first one (walking up the chain using the $\VMK_{i-1}$ field in the VMB). This is the command that actually creates the Linux virtual devices representing the volumes, under \texttt{/dev/mapper}: the names are generated algorithmically. Notice that these virtual devices are not automatically mounted, it is up to the user to mount them and format them with a filesystem of choice when required.

Finally, the \shufflecake \sflcclose command takes a device path as input, and closes all the volumes open on that device.

\paragraph{Additional functionalities.} In addition to standard features such as command-line usage help and printing on screen the current version, our implementation also offers two additional functionalities: a \texttt{changepwd} action, which allows the user to change a volume's password as described in \expref{Section}{sec:userland}, and a \texttt{testpwd} action, which tests whether a provided password unlocks a certain volume (and which one) without actually opening that volume. This might be helpful for the scrupulous user who wants to regularly recall the passwords to decoy volumes, as suggested in \expref{Section}{sec:interrogation}.

\subsection{Space Utilisation}

A few factors influence the disk and RAM space efficiency of Shufflecake, i.e., what part of the storage contains actual data coming from the upper layer, and what part contains metadata, or is otherwise wasted. Overall, with a sensible choice of the parameters, and with reasonable assumptions about the behaviour of the upper layer, we can attain a very low space overhead.

For our `Legacy' implementation, we fixed the block size \blocksize to 4096 bytes, so as to better amortise the per-block space overhead determined by the IVs. We chose $\Sl=256$, and $\maxvols=15$. Since we use AES-CTR-256 as the underlying encryption scheme, we need 16-byte IVs. This led to a choice of $\deltas = 1$, and hence $\Sp=257$: a single 4-KiB IV block (containing 256 IVs) encrypts a 1-MiB slice. To provide a numerical summary of the space utilisation, we observe that in the case of a 1 TiB disk, the resulting theoretical maximum utilisable space is 1019.91 GiB, equal to more than 99.6\% of the physical storage space.

\paragraph{Headers.} With the above parameters, the total size of a volume header is around $\Shdr = \frac{\devbsize}{\Sp} \log\frac{\devbsize}{\Sp}$, roughly equal to 4 MiB per volume header, for a 1-TiB disk: about 60 MiB for the total device header size.

\paragraph{IVs.} As previously discussed, in Shufflecake `Legacy' we store IVs on-disk. With the concrete choice of parameters of our implementation, we have 16-byte IVs encrypting 4096-byte blocks (256 times as much); therefore, we only use $\frac{1}{257}$ ($<$ 0.4$\%$) of the physical data section to store IVs.

\paragraph{Internal fragmentation of slices.} Internal fragmentation is a frequent problem in space allocation, and it is particularly well known and studied in file systems theory. For performance reasons, the block layer only works with the block granularity; the file system, therefore, has to allocate a whole block even if it needs less space to, e.g., host a file. Internal fragmentation is the problem arising from this ``over-allocation''. On top of this, Shufflecake adds another layer of internal fragmentation through its slice mechanism: when a volume requests a block, we reserve a whole slice of $\Sl$ blocks just for that volume. Moreover, we have no means of communicating this over-allocation to the file system layer, which therefore has no way of adapting its behaviour. Thus, we have to hope that a file system does not jump back and forth too wildly, and that it generally tries to fill a group of slices before requesting a new one.

Luckily, some file systems do exhibit this behaviour. For example, the commonly used ext4 file system defines the concept of a \emph{block group}, i.e., a group consisting of 32768 consecutive blocks (which amounts to 128 MiB for 4096-byte blocks). The block allocator of ext4 tries its hardest to keep related files within the same block group; specifically, whenever possible, it stores all \emph{inodes} of a directory in the same block group as the directory; also, it stores all blocks of a file in the same block group as its inode~\cite{ext4}. This feature plays nicely with the value of $\Sl=256$ we chose: a block group encompasses a whole number of slices, which will therefore not be too fragmented in the long run.

\paragraph{Releasing unused slices.} Our implementation currently does not have a way to reclaim physical slices that were assigned to some volume but are no longer used, when all of the blocks within the corresponding logical slice have been deallocated by the file system. We discuss this in~\expref{Section}{sec:reclaimslice}. We note, however, that deallocation of slices can occur more or less frequently depending on the filesystem in use: filesystems with good contiguity will tend to free up consecutive blocks (and hence whole slices) as the data is moved or erased, while filesystems with higher granularity might `leave blocks around' more often. We cannot release a slice until all the physical blocks therein are freed up. Therefore, the efficiency of any slice-releasing mechanism must be evaluated carefully.

\subsection{Benchmarks}\label{sec:benchmarks}

We tested our `Legacy' implementation looking at I/O performance and space efficiency. The test environment was a fresh installation of \osbenchmark running kernel \kernelbenchmark on a laptop equipped with an AMD Ryzen 7 PRO 6850U CPU with Spectre mitigations enabled, 32 GiB 4-channel 6400 MHz DDR5 RAM, and a low-power 1 TiB NVMe Micron MTFDKCD1T0TFK SSD. We tested the amount of slice fragmentation of Shufflecake (\sflcbenchmarkversion), its I/O performance, as well as the I/O performance of other two relevant disk-encryption tools for comparison: \texttt{dm-crypt}/LUKS (\luksbenchmarkversion) and VeraCrypt (\veracryptbenchmarkversion). All the tests were performed sequentially, on a physical primary SSD partition of size 8 GB, using the ext4 filesystem (which is the one most relevant for Shufflecake's envisioned final use case). In the case of Shufflecake, we {\sflcinit}iated the partition with two volumes (one decoy and one hidden), and performed all tests on the hidden one. Analogously, in the case of VeraCrypt, we formatted the partition as a standard no-FS VeraCrypt volume, and created a 6.5 GB ext4 volume therein. In order to aid reproducibility, we also included in our implementation a suite of benchmark scripts performing the tests described here.

\subsubsection{Fragmentation} In order to evaluate the fragmentation caused by Shufflecake's allocation of slices, we filled the ext4 filesystem with incrementally large amount of random files and directories up to saturating the space, and at every step we measured the space given by the increasing number of slices allocated by \dmsflc for the hidden volume VS the total amount of data written therein. We define \emph{space efficiency} as the ratio between real data written on disk and slice-allocated space ($0$ = bad, $1$ = good).

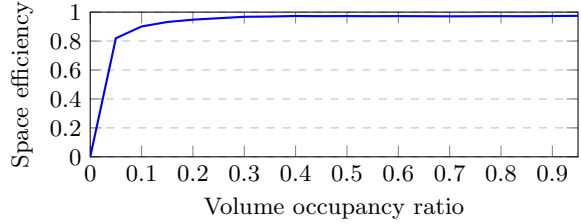
\begin{figure}[ht]
\centering
\begin{tikzpicture}
\pgfplotsset{compat=1.16, width=8cm, height=3.5cm}
\begin{axis}[
    xlabel={Volume occupancy ratio},
    ylabel={Space efficiency},
    xmin=0, xmax=0.95,
    ymin=0, ymax=1.0,
    xtick={0,.1,.2,.3,.4,.5,.6,.7,.8,.9},
    ytick={0,.2,.4,.6,.8,1.0},
    legend pos=north west,
    ymajorgrids=true,
    grid style=dashed,
]
    
\addplot[
    color=blue,
    mark=dot,
    thick,
] coordinates {
(0,0)
(0.05,0.8197)
(0.1,0.9009)
(0.15,0.9317)
(0.2,0.9479)
(0.25,0.9579)
(0.3,0.9677)
(0.35,0.9695)
(0.4,0.9732)
(0.45,0.9719)
(0.5,0.9728)
(0.55,0.9717)
(0.6,0.9724)
(0.65,0.9716)
(0.7,0.9709)
(0.75,0.9715)
(0.8,0.9721)
(0.85,0.9714)
(0.9,0.9730)
(0.95,0.9744)
};

\end{axis}
\end{tikzpicture}
\caption{Shufflecake space efficiency as the ext4 filesystem fills up.}
\label{table:benchfrag}
\end{figure}

The results are shown in~\expref{Figure}{table:benchfrag}. As we can see, even when the disk is initially empty, some slices are immediately allocated for the ext4 journal and metadata. However, as data is written on disk, the effect of fragmentation quickly disappears: already at 10\% of data capacity the space efficiency is above 90\%, and at 25\% of data written it reaches 95\%. We conclude that the slicing algorithm of Shufflecake behaves very well in our simulated random usage pattern, at least with the ext4 filesystem, and slice fragmentation can be considered negligible.

\subsubsection{I/O and bandwidth.} For testing the I/O performance of (`Legacy') Shufflecake against \texttt{dm-crypt}/LUKS and VeraCrypt, we used the \texttt{fio} benchmarking tool, which can flexibly measure various metrics. For each of the three disk encryption tools, we performed both random and sequential read/write operations with large amount of data on the filesystem. We fixed the same parameters for all tests, such as a queue of 32 operations and a block size of 4 kiB, which are commonly recommended to evaluate real-world performance of disks. Under these conditions, we found no observable difference in metrics between IOPs (I/O operations per second) and bandwidth (expressed in MB/s), hence we report the results looking at the bandwidth only.

\begin{table}[ht]
\newlength{\oldabovecaptionskip}
\setlength{\oldabovecaptionskip}{\abovecaptionskip}
\setlength{\abovecaptionskip}{10pt}
\centering
\begin{tabular}{|c|c|c|c|}
\hline
 & ~~~~Shufflecake~~~~ & ~~~~\texttt{dm-crypt}/LUKS~~~~ & ~~~~VeraCrypt~~~~ \\ 
\hline
random write & 26.77 & 38.43 & 39.07 \\
\hline
random read & 26.78 & 38.44 & 39.09 \\ 
\hline
sequential write & 176.87 & 247.14 & 247.75 \\
\hline
sequential read & 177.10 & 247.43 & 248.04 \\
\hline
\end{tabular}
\caption{I/O performance (in MB/s) of Shufflecake, dm-crypt/LUKS, and VeraCrypt (higher = better).}
\label{table:benchall}
\setlength{\abovecaptionskip}{\oldabovecaptionskip}
\end{table}

\vspace{-0.5cm}

The results are shown in~\expref{Table}{table:benchall}. As we can see, Shufflecake incurs in an I/O slowdown of roughly 30\% compared to the other tested tools. We believe this overhead to be acceptable in daily use.

\paragraph{Comparison with woORAMs.} The comparison with some popular ORAM-based PD solutions more convincingly shows the real-world efficiency advantages offered by Shufflecake. Of course it has to be stressed that these ORAM-based solutions aim at achieving PD in a more strict scenario than the single-snapshot security offered by the current version of Shufflecake. We have just seen how Shufflecake achieves a slowdown of roughly 30\% I/O throughput over \texttt{dm-crypt} and uses almost all space available. On the other hand, HiVE~\cite{hive} has a heavy 200x I/O overhead and wastes 50\% of the disk space. DetWoORAM~\cite{detwoORAM} has an overhead of 2.5x for {\nread}s and 10x-14x for {\nwrite}s, and wastes 75\% of space.


\section{Conclusions and Future Directions}\label{sec:conclusion}

We have seen how Shufflecake represents a usable PD scheme with many operational advantages over solutions like TrueCrypt. We released it as an open source tool in the hope of building trust and adoption in the community, and possibly encouraging contribution to future work. In fact, many possibilities for further improvement exist. We are going to mention some in this section.

\subsection{Crash Consistency}\label{sec:crashconsistency}

As it is right now, the main obstacle to reach maturity and adoption for daily use is that (`Legacy') Shufflecake is not crash-consistent. This means that if the program crashes during operation with one or more open volumes, data corruption is possible, because some volume state changes happen in-RAM and are cached for some time before being written on disk. This is a problem for (`Legacy') Shufflecake, because if a crash occurs between a write of encrypted data to disk and its associated IV, the encryption becomes unrecoverable. The situation is different for solutions like LUKS or TrueCrypt, which use the XTS mode of operation and are therefore immune to this problem because they do not use explicit random per-block IVs. To fix this while keeping the property of block re-randomisation, we should make the individual logical write requests atomic, i.e. we should mask the fact that they map to several physical requests (which need to be assumed atomic): if a crash happens at any point between two of these physical requests, the old content of the logical block should still be recoverable, the disk should not be left in a ``limbo'' state that does not correspond to any of the logical contents written by the upper layer. We discuss here some ideas for future improvements to address these concerns.

Shufflecake incurs crash-inconsistencies when a crash happens in the time window between the update of a data block and the update of the corresponding IV block. As was discussed, Shufflecake `Legacy' adopts a write-on-flush approach for the IV cache, whereas data blocks are immediately written to disk, encrypted with the new IV (which is not immediately persisted on-disk); therefore, the disk is in a ``vulnerable'' (inconsistency-prone) state whenever the upper layer has written on a file and has not yet synced it: this is, reasonably, a large fraction of the total operating time. To solve this, it would be necessary to make the IV cache write-through; the performance impact of such a choice has not been evaluated but it would probably be heavy. It would not be sufficient, anyway, because it would only reduce the ``vulnerability window'' between the update of a data block and that of the IV block, it would not eliminate it.

The final solution would be to also duplicate each IV block into a circular log of length 2: the update of the IV block synchronously precedes the update of the data block, and overwrites the older of the two versions; this way, if the crash happens right afterwards, and the data block is not updated, it is still decryptable because the corresponding IV block has not been touched. Disambiguation (i.e., deciding which of the two versions of the IV to use) would be based on an additional MAC on the data block (stored alongside the IV); this would only be needed when the block is read for the first time since volume opening: afterwards, the state can be kept in RAM (it is just one bit for each slice).

An alternative solution, which wastes more disk space but we believe to be overall better, would be to store the IV alongside the data block itself, so that the two updates can be merged into a single physical request. We believe that, in addition to mitigate the issue of crash inconsistency, this approach would probably lead to better I/O performance, as it would not need separate operations for writing IV blocks during data writes. The minimum addressable unit of disk storage space, at least on Linux systems, is usually the 512-byte sector. Therefore, the least wasteful option is to map a logical 4096-byte block (8 sectors, as was already the case for our implementation) onto 9 consecutive physical sectors: the first one contains the IV, the other ones constitute the data block. This would lead to a waste of disk space (fraction of disk not used for upper-layer data) equal to $\frac{1}{9} = 11.1\%$. Since an IV only occupies 16 bytes, much of the first sector would be left unused; we could use the rest of the free space to also contain additional useful data, for example a MAC to detect IV alterations, and a ``reverse map'', indicating which logical block $\Blog$ of which volume $\V_i$ that physical block corresponds to (this information would of course be encrypted). Additionally, we could build an even more fault-resilient system by again having a circular log of two IVs in the first sector (each accompanied by the corresponding MAC), plus 16 MACs, one for each IV and for each of the 8 data sectors. This would allow us to disambiguate with the sector granularity, in case the underlying disk can ensure atomicity of the sector writes, but not of the physical requests writing on several adjacent sectors.

\subsection{Multi-Snapshot Security}\label{sec:multisnapshot}

The way it has been presented so far, Shufflecake `Legacy' is completely vulnerable to multi-snapshot attacks in the exact same way as TrueCrypt (and its successor, VeraCrypt): when the adversary sees ``empty'' slices change across snapshots, the only possible explanation is that there still is a hidden volume whose password has not been provided. We have already argued in~\expref{Section}{sec:previouswork} why, in practice, this level of security might be enough in most cases, and also why we believe current woORAM-based solutions base their promise of stronger security on somewhat hard to justify assumptions. We already know~\cite{sok} that achieving ``complete'' multi-snapshot security requires the use of woORAMs, which have serious performance drawbacks. Regardless, as mentioned in \expref{Section}{sec:sflc}, we designed Shufflecake with the idea of being able to add features that might help to reach an unproven, ``operational'' level of multi-snapshot security. In this section, we explore the possibility to achieve this goal through some separate, ``orthogonal'' pattern obfuscation procedure that operates independently of the main scheme and does not interfere with its single-snapshot security. We present three high-level ideas to accomplish this. They are not currently implemented, nor are they precisely specified from a conceptual point of view; instead, they are left as pointers for future research, since this area is greatly under-studied.

\subsubsection{Security Notion Revisited.} Let us first clarify what we mean by ``operational'' multi-snapshot security. The rationale is the hope that even though the distinguishing advantage of the adversary in the multi-snapshot game is not negligible (and so the scheme is not cryptographically secure in the strict sense), it is still low enough for an investigation to be inconclusive against an adversary in practice. In other words, we argue that \emph{legal} or anyway \emph{operational} security, in this context, is not the same as \emph{cryptographic} security in the theoretical sense.

Physical slices have known boundaries, so the adversary can compare the disk snapshots she has on a slice basis. This amounts to inspecting the subsequent changes, or \emph{diffs}, each physical slice goes through across snapshots. Recall that a physical slice is essentially an array of $\Sp$ blocks; when one of these blocks changes, it can either be because of the re-encryption of a data block (with a different IV, and possibly a different content), or because of the re-randomisation of an empty block: the IND-CPA security of the encryption scheme guarantees the indistinguishability of these two situations. This means that no hint on the nature of a block (i.e., whether it is a data or an empty block) is leaked by its encrypted content or by the history of its encrypted contents. Therefore, when comparing two snapshots of a physical slice, the only information that an adversary gets is \emph{which} of the $\Sp$ blocks have changed: \emph{how} they have changed is completely inconsequential and uninformative. In other words, the diff between two snapshots of a physical slice boils down to a bitmask of $\Sp$ bits, representing which blocks have changed and which remained the same.

The task of the adversary then becomes to distinguish between ``data slices'' (i.e., belonging to some volume) and ``free slices'' (not mapped to any volume) based on a sequence of such diffs for each slice. The point is that, after unlocking the first $\Nvol - $ volumes, the rest of the space may or may not contain another volume (there may or may not be some more data slices among the free slices): if we want the adversary to be incapable of distinguishing between the two cases, we need the diffs of data slices and free slices to look the same. Once we frame the problem in this way, we can rephrase the weakness of Shufflecake as follows: the diff bitmask of a free slice is always all-zeros, and as such is clearly distinguishable from that of a data slice, which might have some bits set to $1$.

Our task then becomes to ``obfuscate'' the changes that occurred in the data slices (especially those belonging to $\V_{\Nvol-1}$) by artificially creating a non-zero diff bitmask in the free slices, through a re-randomisation of some selected blocks. This way, a user will hopefully be able to claim that all the changes happened to the ``empty'' blocks are due to this obfuscation procedure, and not to the existence of $V_{\Nvol-1}$. Nothing prevents us, of course, from also touching the data slices during the obfuscation procedure, if that helps making the diff bitmasks look more alike. It is to be noted, however, that if a block was modified by the upper layer during the normal operational phase, the corresponding bit will be set to 1 in the diff bitmask and there is no way for the obfuscation procedure to ``undo'' that change: when we touch a data slice, we cannot turn the $1$s of the diff bitmask into $0$s. Instead, we can turn some $0$s into 1s by simply re-encrypting the same content of a block with a different IV.

\subsubsection{Trivial Random Refresh.} A very simple first idea for an obfuscation procedure is to take all physical blocks belonging to free slices, and re-randomise them all, independently at random, each with probability $p_1$. This operation could be either performed upon volume \sflcclose or, for better resilience, spread across the normal operations. It is the easiest way to achieve a non-zero diff bitmask for free slices, but it is definitely too crude to work: nothing guarantees that the diff bitmasks of data slices will ``look random'' like the ones artificially generated for free slices. Also, it might very well be the case that many data slices do not change across two snapshots, in which case it becomes very easy to tell them apart from free slices, which often have a non-zero diff bitmask.

Such consideration suggests a refinement: besides re-randomising some blocks in the free slices, we could re-encrypt (with a different IV but same plaintext) some blocks in the data slices of existing volumes, again independently at random, each with probability $p_2$. This way, we also randomise the diff bitmasks of the data slices, making them more similar to those generated for free slices.

\paragraph{Insecurity with many snapshots.} The procedure we just illustrated could well succeed, for a suitable choice of $p_1$ and $p_2$, in rendering the diff bitmasks of data slices and free slices roughly indistinguishable, but only if the adversary gets just \emph{one} diff for each slice, i.e., if she only gets two snapshots. This is because we are essentially playing a hopeless game: roughly speaking, we are aiming at making signal+noise (the diffs of a data slice) indistinguishable from noise alone (the diff of a free slice). As discussed, we cannot turn the $1$s of the diff bitmasks of data slices into $0$s: the ``signal'' given by which blocks were modified by the upper layer stays there, we can only hope to bury it in enough noise by turning some $0$s into $1$s through re-encryption. However, with enough snapshots, the signal will eventually emerge. Imagine, for instance, that there is one particular block in a data slice that is very often modified (maybe it contains some sort of file system index): the corresponding bit in the diff bitmask will often be set to $1$, which would be hard to justify through the obfuscation procedure, which only hits one given block with probability $p_1$ each time.

\subsubsection{Subsampling.} The previous discussion teaches us a valuable lesson: in our setting, we cannot hope to disguise the accesses performed by the upper layer as random noise, as is commonly the case for ORAMs. The only option we have left is to take the opposite approach: let us make the diffs of free slices look like they were \emph{also} generated by some file system workload. This way, we are still making the diffs for the two kinds of slices similar, but we are not trying to erase the signal from existing data slices, which we cannot. Instead, we ``copy this signal'' onto the free slices, so that the adversary will \emph{always} see this signal, even if the last $\Nvol$-th volume has been surrendered. A tedious, convoluted, and yet very imprecise way of doing it would be for us to sit down, study the access patterns resulting from typical file system workloads, model them as a probability distribution, and hardcode that into the obfuscation procedure. Probably, a better idea would be to use a ML approach and let a daemon run in the background to adaptively learn and simulate such distribution.

Instead, a simpler method that is likely to capture the patterns we want to imitate is to have the scheme itself ``learn'' them online, by simply subsampling the stream of incoming logical requests. More concretely: for each incoming logical \SflcWrite request, we ``imitate'' it with probability $p$. Imitating a request means ``learning'' that the affected logical block at position \Blog is likely to be written by file system workloads, and copy this signal onto a free slice: we retain the offset $(\Blog \mod \Sl)$ of the block within the slice, we choose a ``target'' free slice, and we re-randomise the block with the same offset within the target slice. This approach guarantees that, if a particular block is often updated by the upper layer, then we are very likely to catch this signal and correctly carry it over to a free slice. Note, however, that we need to choose the target free slice deterministically from some context information. Only in this way can we replicate the signal consistently across snapshots, always onto the same free slice; also, this allows us to correctly capture and copy a signal in case it consists of not just one, but several blocks in a slice being frequently updated.

\paragraph{Counting attacks.} If we assume that the obfuscation procedure just described really succeeds in making the empty space look like it is occupied, we are still left with one problem. The special blocks that are often updated by the upper layer leave a very clear trace in the snapshot history, since their bits in the diff bitmasks are almost always set to 1. If we have $\Nvol$ volumes, the obfuscation procedure will generate $\Nvol$ such clear traces in the free space. Therefore, a simple attack would consist in counting these traces, and checking whether there are as many of them as there are disclosed volumes. To thwart this attack, we can rework the obfuscation procedure in such a way that the device's data section always looks like it's hosting $\maxvols$ volumes. We can ideally assign the free slices to $\maxvols - \Nvol$ pairwise-disjoint sets, each one representing a ``fake'' volume, and have the obfuscation procedure be aware of this partitioning when choosing the target free slice, so as to really simulate $\maxvols - \Nvol$ volumes with the imitated logical \SflcWrite requests.

\subsubsection{Ghost File System.} The above procedure may already offer good protection, although it might be non-trivial to translate the rough idea of ``being aware of the partitioning into $\maxvols - \Nvol$ fake volumes'' into a concrete algorithm for choosing a target free slice when we imitate a \SflcWrite request. A valid, if ``exotic'', alternative, would be to actually create $\maxvols - \Nvol$ additional ``ghost'' Shufflecake volumes on the device, that behave in the exact same way as regular volumes, except that they do give up their slices when needed. On these volumes, a separate component (a daemon) could mount a file system and perform some typical sequences of accesses. The advantage of this solution is that, by definition, it will always look like there are $\maxvols$ volumes on the device. However, when slices are reclaimed from ghost volumes, their file systems might suddenly get corrupted and start complaining quite loudly, thus impairing the practical usability of the system. A daemon operating on these ghost filesystems might therefore need to be aware of the new slice allocation requests from ``real'' volumes, and move or remap the ghost slices accordingly at runtime.

\subsection{Shufflecake ``Lite''}\label{sec:sflclite}

As we have seen in~\expref{Section}{sec:crashconsistency}, crash inconsistency is a serious problem of the current embodiment (`Legacy') of Shufflecake, caused by the use of the CTR mode of encryption which, in turn, is necessary to achieve block re-randomisation. And we have seen in~\expref{Section}{sec:multisnapshot} how this block re-randomisation is an essential ingredient for achieving some form of multi-snapshot security. However, currently no multi-snapshot security measure is implemented in Shufflecake `Legacy', or even formally defined. We have already seen in~\expref{Section}{sec:benchmarks} how the current CTR design brings some (minimal) performance hit, and how this performance hit will likely be exacerbated both by the discussed mechanisms for achieving crash consistency and by the proposed ideas for multi-snapshot security. Furthermore, we have argued in~\expref{Section}{sec:previouswork} how single-snapshot security might already offer a good enough security margin in many scenarios.

This brings us the following idea: proposing a \emph{``Lite''} mode of Shufflecake which sacrifices the feature of block re-randomisation (with all its pros and cons) and employs the XTS mode of operation instead of CTR for encryption, just like TrueCrypt. This would give up every hope of achieving any form of security better than single-snapshot, but it would bring the following advantages:
\begin{itemize}
\item It would avoid the need of writing IVs on disk, therefore avoiding completely the (minimal) space waste and I/O slowdown measured in~\expref{Section}{sec:benchmarks}.
\item It would be \emph{natively} crash-consistent, so all the countermeasures discussed in~\expref{Section}{sec:crashconsistency} would be unnecessary.
\item Compared to TrueCrypt/VeraCrypt, it would still offer huge operational advantages: providing ``real'' plausible deniability by offering many nested layers of secret volumes, unlocking a whole hierarchy of volumes with a single password, and being filesystem-agnostic.
\end{itemize}

The idea would be eventually to provide both modes for Shufflecake, ``Lite'' and ``Full'' (the latter including both crash consistency and partial multi-snapshot security), and let the user choose which one is desired during the \sflcinit operation. In practice, one of the two would be the default mode and the other one could be selected as optional (while the `Legacy' mode would eventually be deprecated), but deciding which of the two shall be the default will require a careful security and usability study. We leave this decision for future work, after proper confrontation with representatives of the envisioned final user demographics.

\subsection{Corruption Resistance}\label{sec:corruptionresistance}

As discussed in \expref{Section}{sec:opmod}, writing data to decoy volumes while not all the hidden volumes are open entails a risk of corrupting the hidden volumes. There cannot be perfect mitigation to this problem, save for frequent backups. However, it could be possible to reduce this risk of corruption by using some form of \emph{error-correction} on the unopened volumes (hence sacrificing some space), and then trying to recover the volume if corruption happens. 
We tested positively this idea using RAID\cite{raid}, namely by partitioning a Shufflecake hidden volume into different equal-sized logical partitions, and using them to assemble a RAID device with redundancy. Other methods might be more suitable, such as \emph{alpha entanglement codes}\cite{alphacodes} or other error-correcting codes.

At present, we do not have plans of including corruption-recovery capabilities within Shufflecake itself. This is first because volume corruption is an event which, albeit undesirable, is a consequence of unsafe operations rather than a necessary scenario to be covered. And, second, because we do not want to impose on the user a default choice of mitigation tools (e.g., RAID vs alpha entanglement codes vs something else). However, what we plan of doing is facilitating the integration of such corruption-recovery tools with Shufflecake, by providing the necessary technical machinery to make them work efficiently, and by providing an API interface for communication between Shufflecake, the tools, and the OS.

\subsubsection{Improving resolution of corrupted slice assignments.} We have already seen in~\expref{Section}{sec:sflcdesign} how Shufflecake tries to resolve volume corruption by reassigning new slices to disambiguate collisions in the slice maps. We have also seen how this mechanism has the drawback that corruption of a single block of a lower-order volume at the logical level makes the whole slice content unrecoverable to the higher-order volume, because the newly assigned slice contains garbage data. One possible improvement in this respect is to actually \emph{clone} the corrupted slice instead, so to make (potentially most of) its original content available again to the higher-order volume, thereby greatly improving the effectiveness of external corruption mitigation tools. This has to be done carefully, though: We cannot simply take a fresh available PSI and copy the corrupted physical slice at that position, for two reasons.

The first one is that this would break plausible deniability, not only in the multi-snapshot setting, but also if an adversary gets a single snapshot of the device after a corruption recovery attempt happened by slice cloning as just presented. The reason is that the adversary would see two physical slices equal one-to-one in content, which is clearly unexplainable without assuming that such recovery happened - and hence hinting at the presence of at least two volumes.

The second reason is that the blocks in the cloned slice can only be decrypted (by some of the conflicting corrupted volumes) if the correct IVs are also recovered at the new position. This would indeed happen in the CTR mode of operation we use, because IVs are written explicitly as part of the slice, and would therefore also be copied to the new location, but it would not happen, e.g., in the XTS mode of operation, where IVs depend on the position within the device. This means that such a proposed solution would not be compatible with the `Lite' mode of operation of Shufflecake described in~\expref{Section}{sec:sflclite} above.

The right approach is therefore to clone the \emph{plaintext content} of the slice, or at least as much as possible of it, but re-encrypting it (re-randomising it with new IVs) in the new physical slice. This should happen either explicitly (by sampling and storing new fresh IVs on the slice, in the case of CTR mode) or implicitly (by deriving new IVs from the new physical address, in the case of XTS mode). The encryption key to be used, both for decryption and re-encryption, should be that of the higher-order corrupted volume (the one whose position map is being modified). The corrupted blocks within the slice will still be unrecoverable to this volume (but remain available at the original physical position to the lower-order volume which last wrote them), while the non-corrupted blocks will be successfully recovered and re-encrypted to the new location.

The resulting modified \HandleCorruption procedure is depicted in~\expref{Algorithm}{alg:handlecorruption2}. This mechanism allows to recover from volume corruption at the block granularity rather than the slice granularity, thereby reducing the overall amount of error on the reconstructed volumes, and increasing the recovery chances of external mitigation tools.

\vspace{-0.5cm}

\begin{algorithm}[H]
\captionsetup{font={footnotesize}}
\caption{$\HandleCorruption(\V_i, \PSI, \LSI)$}\label{alg:handlecorruption2}
\begin{algorithmic}[1]\footnotesize
\State $\PSI^\mathsf{new} = \NewSlice(\V_i, \LSI)$
\State \textbf{for} $j=0, \ldots , \Sl-1$
\State ~~~~ $\Bphy^\mathsf{in} := \PSI \cdot \Sp + \deltas + j$
\State ~~~~ $\IV^\mathsf{in} \leftarrow \LoadIV\left(\Bphy^\mathsf{in}\right)$
\State ~~~~ $\ctxt \leftarrow \bRead\left(\Bphy^\mathsf{in}\right)$
\State ~~~~ $\ptxt := \Decrypt\left(\ctxt, \IV^\mathsf{in}, \VEK_i\right)$
\State ~~~~ $\Bphy^\mathsf{out} := \PSI^\mathsf{new} \cdot \Sp + \deltas + j$
\State ~~~~ $\IV^\mathsf{out} \leftarrow \SampleAndStoreIV\left(\Bphy^\mathsf{out}\right)$
\State ~~~~ $\ctxt := \Encrypt\left(\ptxt, \IV^\mathsf{out}, \VEK_i\right)$
\State ~~~~ $\bWrite\left(\Bphy^\mathsf{out}, \ctxt\right)$
\State \textbf{next} $j$
\State \Return
\end{algorithmic}
\end{algorithm}

\vspace{-0.8cm}

\subsubsection{Corruption mitigation API}. Another possible improvement in integrating corruption-mitigation capabilities in Shufflecake is to provide an interface to communicate to upper layers (e.g., the OS, or third-party error-correction tools) whether a volume corruption happened, and possibly where. This can be done, e.g., by publishing a virtual \texttt{corruption} file on \texttt{/sys/devices/sflc/\$VOLNAME}. By detecting whether this file exists, the OS or other recovery tools can know in advance when to start a recovery process without need of user intervention. Whether a volume is corrupted or not should be detected by \dmsflc during instantiation, as explained above. Then, after the conflicting slice assignment has been resolved, Shufflecake would set a persistent \texttt{corrupted} flag on the affected volume. Whenever a volume is {\sflcopen}ed and such a set flag is read, Shufflecake would create the \texttt{corruption} file as explained above. In case extra information on the corruption event is known, for example the LSI of affected slices (only known the first time their corruption is detected, i.e., during the disambiguation procedure explained above), this information could be published, e.g., in \texttt{/sys/devices/sflc/\$VOLNAME/corruption}. This might be helpful in case certain external recovery tools work better by knowing the location of errors.

After the recovery process has completed successfully, the external tool or OS should communicate it to \dmsflc, for example by writing a new file \texttt{repaired} at the same path. Then, on device \sflcclose, Shufflecake would check if such file exists and, if so, clear the persistent flag on the affected volume.

\subsection{Use of Disk Metadata}\label{sec:metadata}

We have seen in~\expref{Section}{sec:sflcdesign} how there is space for embedding volume-specific metadata in each volume's VMB. Here we discuss a couple of useful ideas on how to employ this extra space.

One option could be to embed a string specifying a user-defined name for the volume. Currently, our implementation assigns volume names procedurally in order to avoid name collisions, but a user might prefer to assign these names statically or with mnemonic IDs, to facilitate scripting etc.

Likewise, one could embed a string specifying a desired \emph{mountpoint} for that volume. Notice in fact that, given the PD requirements, one cannot let these volumes be assigned at static mountpoints in a regular Linux way, e.g. using \texttt{fstab} or \texttt{crypttab}. Rather, the desired mountpoint should be hidden within the context of the volume itself. Then, if the implementation supports it, once the volume is {\sflcopen}ed, it can also be automatically \texttt{mount}ed at a given position.

Another idea could be to embed \emph{virtual quotas}, in order to artificially limit the maximum available size of decoy volumes. As it is now, Shufflecake performs maximum overcommitment on the visible available space of all volumes: each of them will appear as large as the underlying device. This can cause issues if the user (or the OS) mistakenly assumes that that space is actually available, and starts writing too much data on the volumes. In order to mitigate this, metadata could be used to limit the size of the block device seen by the OS. Importantly, this must only hold for decoy volumes, because overcommitment is substantially what allows for PD. Therefore, the correct way to implement this is: \emph{each volume's VMB should specify a virtual quota for the volume below itself in the secrecy hierarchy, but not for itself.} When an $\Nvol$-th volume is {\sflcopen}ed, the virtual quota for all less secret volumes in the hierarchy will be recursively read this way (the lowest volume, e.g. volume $0$, will not have this assigned metadata, or anyway it will be ignored). Then, the $\Nvol$-th volume could be assigned a virtual size equal to the maximum available size on the device, minus the \emph{sum} of the virtual quotas of all other volumes. This way, it will always be impossible to accidentally write too much data on the volume hierarchy, but an adversary will always see the most secret unlocked volume as \emph{the} last one present.

Volume metadata is also the natural right place to store a persistent corruption flag, as explained in~\expref{Section}{sec:corruptionresistance}.

It might also be possible, although probably overkill and exceedingly complex to implement, to have different security or redundancy policies assigned \emph{per-volume} rather than \emph{per-device}, and use the metadata to disambiguate them. For example, it could in theory be possible to have different features in terms of crash consistency (as discussed in~\expref{Section}{sec:crashconsistency}), security (\expref{Section}{sec:multisnapshot}) or corruption resistance (\expref{Section}{sec:corruptionresistance}) assigned to different volumes within the same hierarchy.

Finally, all this metadata could be embedded in raw text, or a more robust and machine-friendly encoding such as JSON could be used.

\subsection{Reclaiming Unused Slices}\label{sec:reclaimslice}

Currently, our implementation of Shufflecake does not have a mechanism for reclaiming slices that are no longer used: once a slice is allocated for a certain volume, it will always belong to that volume, even if the volume's filesystem is emptied of all data. It would be desirable to implement an operation to reassign empty slices to the pool of free available ones, in order to make space allocation across volumes more efficient and limiting the risk of overcommitment.

Clearly, we need some sort of hint from the upper layer in order to trigger this operation. To that effect, we need to intercept the \texttt{trim} requests emitted by the file system. These commands are effectively a third instruction accepted by hard disks, besides \nread and \nwrite; they serve as a way for the file system to indicate to the disk that some sectors no longer contain user data, and so the internal disk controller can avoid copying them over when reshuffling its own internal indirection layer~\cite{trim}. These commands are also vital for the efficiency of disk virtualisation systems, such as Shufflecake, that overcommit the total underlying space and thus need to exploit every occasion to optimise the resource allocation. 

Once we have this mechanism in place, we can design a way to reassign a freshly freed-up physical slice to the pool of available ones at a random position, so that the function \NewSlice (\expref{Algorithm}{alg:NewSlice}) will return it with uniformly random probability when a new slice is required. More concretely: suppose that Shufflecake intercepts an OS signal telling us that a logical slice for volume $\V_i$ at LSI $\LSI$ is now free. We define a function \ReclaimSlice (\expref{Algorithm}{alg:reclaimslice}) which operates on the same structures used by \NewSlice, and also on the slice map of the interested volume. This function clears the occupation bitfield of the reclaimed slice and the entry in the slice map, then moves the PSI at a random position of \prmslices (after \octr) by doing another Fisher-Yates iteration. We use a subfunction \ReverseShuffle which, given as input a PSI, returns the index of \prmslices where this PSI is found, or error if not present. The way to implement this subfunction can vary, e.g. by keeping in-memory a reverse map of \prmslices, or by doing a linear search every time.

\begin{algorithm}[H]
\captionsetup{font={footnotesize}}
\caption{$\ReclaimSlice(\V_i, \LSI)$}\label{alg:reclaimslice}
\begin{algorithmic}[1]\footnotesize
\State $\PSI \leftarrow \SliceMap_i[\LSI] $
\State $\ofld[\PSI] := \free$
\State $\SliceMap_i[\LSI] := \bot$
\State $\ell \from \ReverseShuffle(\PSI)$
\State \textbf{if} $\ell > \octr$ \textbf{then:} \Return \Comment{No need to reshuffle in this case.}
\State $\mathtt{swap}(\prmslices[\ell],\prmslices[\octr])$
\State $j \fromunif \set{\octr, \ldots, \maxslices - 1}$ \Comment{$\maxslices = \lceil \frac{\devbsize-\Shdr}{\Sp} \rceil$ max number of slices.}
\State $\mathtt{swap}(\prmslices[j],\prmslices[\octr])$
\State \textbf{if} $\ofld[\prmslices[\octr]] = \free$ \textbf{then:} $\octr := \octr - 1$
\State \Return
\end{algorithmic}
\end{algorithm}

\subsection{Unbounded Number of Volumes}\label{sec:unbound}

Shufflecake assumes a number \maxvols of possible volumes that can be provided by any device. Even if this limit can be chosen freely by implementations, it would be desirable to have a way for creating unlimited numbers of volumes (subject to space availability) per device. This would not only remove an artificial limitation on the scheme, but would also strengthen its operational security by making any kind of safeword-like technique (as discussed in~\expref{Section}{sec:safeword}) impossible.

For this to be achievable, volume headers should probably not be adjacent and packed at the beginning of the disk. One idea for further investigation would be to embed every header (except the first, `less secret' one, which is still going to be at the beginning) at random positions within the device, and having them linked by the previous volume header through an ad-hoc pointer field which is \emph{always} present, and indistinguishable from random without the correct password. Traversing this list of linked headers, however, presents some challenges. In particular, when the user provides one password on volume instantiation, how do we know whether the password is wrong? And, if not, how do we reach the right header unlocked by that password? It might be possible to devise some complex linking scheme for an arbitrary number of ``bogus'' headers on the device, but in any case the following limitations would apply.

First, bogus headers cannot ``reserve'' an area of the disk, otherwise we would waste too much space. They can be placed at any position during device initialization, but when a new slice allocation falls on their position, that space should be released. One can think of different ways to handle this, for example by dynamically moving out bogus headers to another free position if they are about to be overwritten, or simply accepting the risk of breaking the list at some random point during use (as this would not impact consistency for the ``real'' volumes, and it would still be enough to justify the impossibility of an a-priori generated safeword). In any case, except for this difference in reserving allocation, ``bogus'' and ``real'' headers should either be treated equally, or extra precautions should be adopted to maintain PD.

Second, once the user inserts a password to instantiate a device, we might not be able to tell anymore whether the password unlocks something or it's wrong (e.g., a typo). Instead, depending on the chosen solution, the program might continue to traverse the linked list in search of something to decrypt with that password, until either it finds the right header, or the list is broken (e.g., by a bogus header which was overwritten), in which case we can say the password was wrong. We might even envision that the user should expect to terminate manually the program in case a provided password does not succeed after some time, because any hardcoded timeout in the implementation could nullify this feature by inserting a de-facto artificial limit to the number of possible headers.

A possible way to implement this idea could be the following:
\begin{enumerate}
\item Coalesce DMB and VMBs into unified, per-volume headers.
\item Each header is one slice large.
\item Every header also contains a field with a random value \textsf{nxtptr}.
\item Except for the first one, headers are found at random disk positions that are a (public) function of the previous header's \textsf{nxtptr}.
\item During \sflcinit, in case there is a collision during a \textsf{nxtptr} generation over the location of another pre-existing header, the currently generated \textsf{nxtptr} value is discarded and sampled again, until a suitable one is found by brute-forcing (since the total header size is supposed to be negligible in comparison to the device size, this should be very efficient).
\item All the headers are functionally equivalent and contain the same fields.
\item The first header contains a value that is the KDF's salt, while the same field in other headers is either left unused, or used to re-salt the password-derived key for every header using a (fast) hash function.
\item Shufflecake would allocate slices for the volumes in the usual way, just considering the slices at header locations as permanently occupied.
\end{enumerate}

The above idea might very well work, but it remains to specify an efficient way to embed this way also the slice maps, which might be larger than one slice. Many options are open to evaluation here, from linked lists starting at the header, to multiple branching pointers.

There might be other good ways to implement the possibility of having a virtually unlimited number of volumes, we leave this for future exploration.

\subsection{Hidden Shufflecake OS}\label{sec:hiddenos}

As discussed in~\expref{Section}{sec:limitations}, a PD solution that only provides volumes for data storage will never achieve a satisfying level of operational security due to leakage from the OS and other applications installed therein. In order to solve this issue, it is important that the OS itself is run from within a hidden volume, as it was done with TrueCrypt's concept of hidden OS. The natural evolution for Shufflecake would be to be launched at boot time (e.g., as a GRUB module~\cite{grub}) and boot a whole Linux distribution installed within a volume. Alternatively, an ad-hoc, minimal Shufflecake bootloader could be deployed.

More concretely, eventually Shufflecake could become itself a full PD-focused Linux distribution, where during installation the user is guided in the process of creating volumes and installing other distributions therein in a guided way. For operational efficiency and security, every OS at layer $j$ should be aware of the filesystem and OS in the volume at layers $i < j$ (which is made possible by the hierarchy among Shufflecake volumes). This would also allow a \emph{butler daemon} to run from the currently running OS and operate in the background on lower-hierarchy OSes, e.g. by performing system updates, downloading emails, etc., so that all these decoy systems are kept up-to-date even if the user neglects to use them regularly. This would in turn allow to ease the suspicion of an adversary when surrendering a decoy password.

As an alternative to having a full Linux distribution for every volume, a hypervisor-based solution like Qubes OS~\cite{qubesos} might be used instead. However, in order to validate this approach, further analysis is required to ensure that the hypervisor (which is not designed with PD in mind) does not leak the existence of hidden volumes.


\newpage
\bibliographystyle{splncs04}
\bibliography{bibliography.bib}

\begin{thebibliography}{10}
\providecommand{\url}[1]{\texttt{#1}}
\providecommand{\urlprefix}{URL }
\providecommand{\doi}[1]{https://doi.org/#1}

\bibitem{stego1998}
Anderson, R., Needham, R., Shamir, A.: {The Steganographic File System}. In:
  Lecture Notes in Computer Science (01 2000). \doi{10.1007/3-540-49380-8_6}

\bibitem{guardian.ripa}
Ashtana, A.: {Revealed: British councils used Ripa to secretly spy on public}.
  [Online; accessed 2023-10-04]
  \url{https://www.theguardian.com/world/2016/dec/25/british-councils-used-investigatory-powers-ripa-to-secretly-spy-on-public}
  (2016)

\bibitem{grub}
Babar, Y., Babar, Y.: {GRUB} bootloader. {Hands-on Booting: Learn the Boot
  Process of Linux, Windows, and Unix} pp. 133--181 (2020)

\bibitem{bbc.oliverdrage}
{BBC}: {Man jailed over computer password refusal}. [Online; accessed
  2023-10-04] \url{https://www.bbc.com/news/uk-england-11479831} (2010)

\bibitem{argon2}
Biryukov, A.: Argon2. [Online; accessed 2023-10-04]
  \url{https://www.password-hashing.net/\#argon2} (2013)

\bibitem{hive}
Blass, E.O., Mayberry, T., Noubir, G., Onarlioglu, K.: {Toward Robust Hidden
  Volumes Using Write-Only Oblivious RAM}. In: {Proceedings of the 2014 ACM
  SIGSAC Conference on Computer and Communications Security}. p. 203–214. CCS
  '14, {Association for Computing Machinery}, {New York, NY, USA} (2014).
  \doi{10.1145/2660267.2660313}, \url{https://doi.org/10.1145/2660267.2660313}

\bibitem{guardian.msando}
Burke, J.: {Kenyan election official `tortured and murdered' as fears of
  violence grow}. [Online; accessed 2023-10-04]
  \url{https://www.theguardian.com/world/2017/jul/31/kenyan-election-official-christopher-msando-dead-before-national-vote}
  (2017)

\bibitem{bbc.rabbani}
Casciani, D.: {Why Cage director was guilty of withholding password}. [Online;
  accessed 2023-10-04] \url{https://www.bbc.com/news/uk-41394156} (2017)

\bibitem{sqoram}
Chakraborti, A., Sion, R.: {SqORAM: Read-Optimized Sequential Write-Only
  Oblivious RAM}. {Proceedings on Privacy Enhancing Technologies}
  \textbf{2020},  216 -- 234 (2017),
  \url{https://api.semanticscholar.org/CorpusID:201070384}

\bibitem{sok}
Chen, C., Liang, X., Carbunar, B., Sion, R.: {SoK: Plausibly Deniable Storage}.
  {Proceedings on Privacy Enhancing Technologies}  \textbf{2022},  132--151 (04
  2022). \doi{10.2478/popets-2022-0039}

\bibitem{schneiertc}
Czeskis, A., Hilaire, D.J.S., Koscher, K., Gribble, S.D., Kohno, T., Schneier,
  B.: {Defeating Encrypted and Deniable File Systems: TrueCrypt v5. 1a and the
  Case of the Tattling OS and Applications}. In: {USENIX Summit on Hot Topics
  in Security (HotSec)} (2008)

\bibitem{sha3}
Dworkin, M.: {SHA-3 Standard: Permutation-Based Hash and Extendable-Output
  Functions} (2015-08-04 2015). \doi{https://doi.org/10.6028/NIST.FIPS.202}

\bibitem{fisheryates}
Eberl, M.: {Fisher–Yates shuffle}. {Archive of Formal Proofs}  (September
  2016), \url{https://isa-afp.org/entries/Fisher_Yates.html}, Formal proof
  development

\bibitem{alphacodes}
Estrada-Galiñanes, V., Miller, E., Felber, P., Pâris, J.F.: {Alpha
  Entanglement Codes: Practical Erasure Codes to Archive Data in Unreliable
  Environments}. In: 2018 48th Annual {IEEE/IFIP} International Conference on
  Dependable Systems and Networks ({DSN}). pp. 183--194. {IEEE Computer
  Society}, {Los Alamitos, CA, USA} (jun 2018). \doi{10.1109/DSN.2018.00030},
  \url{https://doi.ieeecomputersociety.org/10.1109/DSN.2018.00030}

\bibitem{luks}
{Fedora Project}: {LUKS}. [Online; accessed 2023-10-04]
  \url{https://docs.fedoraproject.org/en-US/quick-docs/encrypting-drives-using-LUKS/}
  (2022)

\bibitem{sha256}
Frankel, S., Kelly, S.G.: {Using HMAC-SHA-256, HMAC-SHA-384, and HMAC-SHA-512
  with IPsec}. {RFC} 4868 (May 2007). \doi{10.17487/RFC4868},
  \url{https://www.rfc-editor.org/info/rfc4868}

\bibitem{trim}
Frankie, T., Hughes, G., Kreutz-Delgado, K.: {A Mathematical Model of the TRIM
  Command in NAND-Flash SSDs}. In: Proceedings of the 50th Annual Southeast
  Regional Conference. p. 59–64. {ACM-SE} '12, Association for Computing
  Machinery, New York, {NY}, {USA} (2012). \doi{10.1145/2184512.2184527},
  \url{https://doi.org/10.1145/2184512.2184527}

\bibitem{libgcryptargon}
{GNU Project}: {Libgcrypt 1.10.1 released}. [Online; accessed 2023-10-04]
  \url{https://lists.gnu.org/archive/html/info-gnu/2022-03/msg00007.html}
  (2022)

\bibitem{libgcrypt}
{GNU Project}: Libgcrypt. [Online; accessed 2023-10-04]
  \url{https://gnupg.org/software/libgcrypt/index.html} (2023)

\bibitem{aes}
Heron, S.: {Advanced Encryption Standard (AES)}. Network Security
  \textbf{2009}(12),  8--12 (2009).
  \doi{https://doi.org/10.1016/S1353-4858(10)70006-4}

\bibitem{wiki.in_re_boucher}
{In re Boucher}: {In re Boucher --- Wikipedia, The Free Encyclopedia}. [Online;
  accessed 2023-10-04] \url{https://en.wikipedia.org/wiki/In_re_Boucher} (2022)

\bibitem{xts}
{Institute of Electrical and Electronics Engineers}: {IEEE Standard for
  Cryptographic Protection of Data on Block-Oriented Storage Devices}. {IEEE
  Std 1619-2018 (Revision of IEEE Std 1619-2007)} pp. 1--41 (2019).
  \doi{10.1109/IEEESTD.2019.8637988}

\bibitem{deftl}
Jia, S., Xia, L., Chen, B., Liu, P.: {DEFTL: Implementing Plausibly Deniable
  Encryption in Flash Translation Layer}. In: {Proceedings of the 2017 ACM
  SIGSAC Conference on Computer and Communications Security}. p. 2217–2229.
  {CCS} '17, Association for Computing Machinery, New York, {NY}, {USA} (2017).
  \doi{10.1145/3133956.3134011}, \url{https://doi.org/10.1145/3133956.3134011}

\bibitem{katzlindell}
Katz, J., Lindell, Y.: Introduction to Modern Cryptography. Chapman and
  Hall/{CRC} Press (2007)

\bibitem{wiki.key_disc_law}
{Key Disclosure Law}: {Key disclosure law --- Wikipedia, The Free
  Encyclopedia}. [Online; accessed 2023-10-04]
  \url{https://en.wikipedia.org/wiki/Key_disclosure_law} (2022)

\bibitem{yesoramlowerbound}
Larsen, K.G., Nielsen, J.B.: {Yes, There is an Oblivious RAM Lower Bound!} In:
  Shacham, H., Boldyreva, A. (eds.) Advances in Cryptology -- CRYPTO 2018. pp.
  523--542. Springer International Publishing (2018)

\bibitem{ext4}
Linux: {ext4 High Level Design}. [Online; accessed 2023-10-04]
  \url{https://docs.kernel.org/filesystems/ext4/overview.html} (2022)

\bibitem{stegfs}
McDonald, A.D., Kuhn, M.G.: {StegFS: A steganographic file system for Linux}.
  In: International Workshop on Information Hiding. pp. 463--477. Springer
  (1999)

\bibitem{gcm}
McGrew, D., Viega, J.: {The Galois/counter mode of operation (GCM)}.
  {Submission to NIST Modes of Operation Process}  \textbf{20},  0278--0070
  (2004)

\bibitem{bitlocker}
Microsoft: {BitLocker}. [Online; accessed 2023-10-04]
  \url{https://docs.microsoft.com/en-us/windows/security/information-protection/bitlocker/bitlocker-overview}
  (2022)

\bibitem{kernel-crypto}
Mueller, S.: {Kernel Crypto API}. [Online; accessed 2023-10-04]
  \url{https://www.kernel.org/doc/html/v4.16/crypto/index.html} (2022)

\bibitem{oechslin}
Oechslin, P.: {Making a Faster Cryptanalytic Time-Memory Trade-Off.} In: Boneh,
  D. (ed.) {CRYPTO}. Lecture Notes in Computer Science, vol.~2729, pp.
  617--630. Springer (2003),
  \url{http://dblp.uni-trier.de/db/conf/crypto/crypto2003.html#Oechslin03}

\bibitem{raid}
Patterson, D., Gibson, G., Katz, R.: {A case for Redundant Arrays of
  Inexpensive Disks (RAID)}. {ACM SIGMOD Record}  \textbf{17} (07 1988).
  \doi{10.1145/50202.50214}

\bibitem{scrypt}
Percival, C., Josefsson, S.: {The scrypt Password-Based Key Derivation
  Function}. {RFC 7914} (Aug 2016). \doi{10.17487/RFC7914},
  \url{https://www.rfc-editor.org/info/rfc7914}

\bibitem{defy}
Peters, T., Gondree, M., Peterson, Z.: {DEFY: A Deniable, Encrypted File System
  for Log-Structured Storage}. In: Network and Distributed System Security
  Symposium (01 2015). \doi{10.14722/ndss.2015.23078}

\bibitem{tcsatyagraha}
Register, T.: Brazilian banker's crypto baffles {FBI}. [Online; accessed
  2023-10-04]
  \url{https://www.theregister.com/2010/06/28/brazil_banker_crypto_lock_out/}
  (2010)

\bibitem{tcmiranda}
Reuters: {UK asked N.Y. Times to destroy Snowden material}. [Online; accessed
  2023-10-04]
  \url{https://www.reuters.com/article/us-usa-security-snowden-nytimes-idUSBRE97T0RC20130830}
  (2013)

\bibitem{wiki.ripa}
{RIPA}: {Regulation of Investigatory Powers Act 2000 --- Wikipedia, The Free
  Encyclopedia}. [Online; accessed 2023-10-04]
  \url{https://en.wikipedia.org/wiki/Regulation_of_Investigatory_Powers_Act_2000}
  (2022)

\bibitem{detwoORAM}
Roche, D.S., Aviv, A., Choi, S.G., Mayberry, T.: Deterministic, stash-free
  write-only {ORAM}. In: Proceedings of the 2017 {ACM SIGSAC} Conference on
  Computer and Communications Security. pp. 507--521 (2017)

\bibitem{wiki.rubber_hose}
{Rubber-hose cryptanalysis}: {Rubber-hose cryptanalysis --- Wikipedia, The Free
  Encyclopedia}. [Online; accessed 2023-10-04]
  \url{https://en.wikipedia.org/wiki/Rubber-hose_cryptanalysis} (2022)

\bibitem{qubesos}
Rutkowska, J., Wojtczuk, R.: {Qubes OS architecture}. Invisible Things Lab Tech
  Rep  \textbf{54}, ~65 (2010)

\bibitem{blake2}
Saarinen, M.J.O., Aumasson, J.P.: {The BLAKE2 Cryptographic Hash and Message
  Authentication Code (MAC)}. {RFC 7693} (Nov 2015). \doi{10.17487/RFC7693},
  \url{https://www.rfc-editor.org/info/rfc7693}

\bibitem{modesofoperation}
Stallings, W.: {NIST block cipher modes of operation for confidentiality}.
  Cryptologia  \textbf{34}(2),  163--175 (2010)

\bibitem{refugee}
Star, T.: {How a Syrian refugee risked his life to bear witness to atrocities}.
  [Online; accessed 2023-10-04]
  \url{https://www.thestar.com/news/world/2012/03/14/how_a_syrian_refugee_risked_his_life_to_bear_witness_to_atrocities.html}
  (2012)

\bibitem{sflcwebsite}
{The Shufflecake Project}: {The Shufflecake Website}. [Online; accessed
  2023-10-04] \url{https://shufflecake.net} (2022)

\bibitem{truecrypt}
{The TrueCrypt Foundation}: {TrueCrypt Homepage}. [Online; accessed 2023-10-04]
  \url{https://www.truecrypt71a.com/} (2015)

\bibitem{tcjohndoe}
{United States Court of Appeals, Eleventh Circuit}: {UNITED STATES OF AMERICA
  v. JOHN DOE, Nos. 11–12268 \& 11–15421 - D.C. Docket No.
  3:11–mc–00041–MCR–CJK}. [Online; accessed 2023-10-04]
  \url{https://caselaw.findlaw.com/court/us-11th-circuit/1595245.html} (2012)

\bibitem{wiki.us_v_fricosu}
{United States v. Fricosu}: {United States v. Fricosu --- Wikipedia, The Free
  Encyclopedia}. [Online; accessed 2023-10-04]
  \url{https://en.wikipedia.org/wiki/United_States_v._Fricosu} (2022)

\bibitem{veracrypt}
VeraCrypt: {VeraCrypt Homepage}. [Online; accessed 2023-10-04]
  \url{https://www.veracrypt.fr/en/Home.html} (2022)

\bibitem{bbc.campaigners}
Ward, M.: {Campaigners hit by decryption law}. [Online; accessed 2023-10-04]
  \url{http://news.bbc.co.uk/2/hi/technology/7102180.stm} (2007)

\end{thebibliography}


\appendix

\newpage
\section{Changelog}

\subsection*{2024-08-23}

Second revision of this document. Release reference software Shufflecake v0.4.5.

\begin{itemize}
\item Fix a number of typos.
\item Clarify that this document describes the `Legacy' architecture of Shufflecake, hinting at future `Lite' and `Full' versions of the Shufflecake scheme.
\item Clarify wording in~\expref{Section}{sec:introduction} to answer common questions.
\item Mention that TrueCrypt and VeraCrypt also support NTFS container volumes under Windows, not only FAT, albeit with serious limitations. Thanks to Long Chongyu for pointing that out.
\end{itemize}

\subsection*{2023-12-07}

First revision of this document. Release reference software Shufflecake v0.4.3.

\begin{itemize}
\item Add this Changelog in appendix.
\item Add Key and Glossary in appendix.
\item Change and restyle symbols and variable names in order to be closer to the reference code implementation.
\item Switch from math-style array indices beginning from $1$ to coding-style array indices beginning from $0$.
\item Add paragraph \emph{``Handling slice conflicts''} in~\expref{Section}{sec:sflcdesign}.
\item Add paragraph \emph{``Device instantiation''} and full algorithm for device instantiation in~\expref{Section}{sec:sflcdesign}.
\item Expand~\expref{Section}{sec:corruptionresistance} with discussion on block-level corruption recovery (and related~\expref{Algorithm}{alg:handlecorruption2}) and on corruption mitigation API.
\end{itemize}

\subsection*{2023-10-06}

First version of this document. Release reference software Shufflecake v0.4.1.

\newpage
\section{Key and Glossary}

\begin{flalign*}
\block & \quad \text{Block of data (typically a 4096 bytes string.)} & \\
\blocksize & \quad \text{Blocksize, in bits (typically 4096 bytes = 32768 bits).} & \\
\dvc & \quad \text{Block device (partition, USB drive, file-backed loop device, SD card, etc).} & \\
\devbsize & \quad \text{Size (in blocks) of a given device.} & \\
\V_i & \quad \text{Logical Shufflecake volume of index $i$ (the higher, the more secret).} & \\
\Nvol & \quad \text{Number of volumes created (from $\V_0$ to $\V_{\Nvol - 1}$).} & \\
\maxvols & \quad \text{Maximum number of volumes per device (15 in our implementation).} & \\
\pwd & \quad \text{User-provided password.} & \\
\keysize & \quad \text{Bitsize of a generic encryption key.} & \\
\KEK & \quad \text{Key Encryption Key, derived from password, used to encrypt a DMB cell.} & \\
\DMB & \quad \text{Device Master Block. Contains \maxvols DMB cells.} & \\
\cell & \quad \text{DMB cell. Contains (encrypted with a KEK) the VMK for a volume.} & \\
\VMK & \quad \text{Volume Master Key, used to encrypt a VMB.} & \\
\VMB & \quad \text{Volume Master Block. Contains information on the associated volume.} & \\
\Shdr & \quad \text{Size (in blocks) of a device header (DMB + VMBs).} & \\
\metadata & \quad \text{metadata associated to a volume. Contained in the volume's VMB.} & \\
\VEK & \quad \text{Volume Encryption Key, encrypts a volume's position map and data slices.} & \\
\Sl & \quad \text{Size (in blocks) of a logical slice.} & \\
\Sp & \quad \text{Size (in blocks) of a physical slice.} & \\
\deltas & \quad \text{Difference (in blocks) between \Sp and \Sl, used to store IVs for CTR mode.} & \\
\PSI & \quad \text{Physical slice index (PSI), i.e., position of a physical slice within a device.} & \\
\LSI & \quad \text{Logical slice index (LSI), i.e., position of a logical slice within a volume.} & \\
\SliceMap_i & \quad \text{Slice map of volume $\V_i$.} & \\
\numslices & \quad \text{Number of logical slices available across all volumes.} & \\
\maxslices & \quad \text{Maximum supported number of slices for a given device.} & \\
\Blog & \quad \text{Logical block address, i.e., index of a block within a volume.} & \\
\Bphy & \quad \text{Physical block address, i.e., index of a block within a device.} & \\
\prmslices & \quad \text{Array of permuted PSIs of a device (reshuffled at every instantiation).} & \\
\ofld & \quad \text{Array of occupation flags, with PSIs as indices and \free/\occupied values.} & \\
\octr & \quad \text{Occupation counter for \prmslices: the first \octr elements are \occupied.} & \\
\end{flalign*}

\end{document}